\documentclass[useAMS,usenatbib]{mn2e}
\usepackage{graphicx}    
\usepackage{subfigure}
\usepackage[latin1]{inputenc}
\usepackage{amsmath}
\usepackage{amsfonts}
\usepackage{amssymb}
\usepackage{color}
\usepackage{pdflscape}

\def \fps@figure{htbp}
\makeatother

\DeclareGraphicsExtensions{.ps,.pdf,.png}
\makeatletter

\def \sersic  {S\'{e}rsic}
\def \reunit  {h$^{-1}$kpc}
\def \kms     {km/s}
\def \reff    {\textit{RFF}}

\def \re      {$R_{\rm e}$}
\def \rkron   {$R_{\rm kron}$}
\def \rffs    {$RFF_{1{\rm c}}$}

\def \chisq   {$\chi^2_\nu$}

\def \vdvir   {$\sigma_{r200}$}

\def \rex      {R_{\rm e}}

\def \rffsx    {RFF_{1{\rm c}}}
\def \rffbdx   {RFF_{2{\rm c}}}

\def \chisx    {\chi^2_{\nu,1c}}
\def \chibdx   {\chi^2_{\nu,2c}}

\def \bestb   {optimal border}
\def \fscore  {\textit{F}-score}

\begin{document}

\title[The Link Between Morphology and Structure of BCGs]{The Link Between Morphology and Structure of Brightest Cluster Galaxies: Automatic Identification of cDs}
\author[Zhao et al.]{Dongyao Zhao$^{1}$\thanks{E-mail: : \texttt{ppxdz1@nottingham.ac.uk}}, 
Alfonso Arag\'{o}n-Salamanca$^{1}$\thanks{E-mail: : \texttt{alfonso.aragon@nottingham.ac.uk}}, 
Christopher~J.~Conselice$^{1}$\thanks{E-mail: : \texttt{christopher.conselice@nottingham.ac.uk}}\\
\footnotemark[0]\\$^{1}$School of Physics and Astronomy, The University of Nottingham, University Park, Nottingham, NG7 2RD, UK}

\date{Accepted 2015 January 24. Received 2015 January 20; in original form 2014 December 12}
\pagerange{\pageref{firstpage}--\pageref{lastpage}} \pubyear{2014}
\maketitle

\label{firstpage}

\begin{abstract}

We study a large sample of $625$ low-redshift brightest cluster galaxies (BCGs) and link their morphologies to their structural properties. We derive visual morphologies and find that $\sim57$\% of the BCGs are cD galaxies, $\sim13$\% are ellipticals, and $\sim21$\% belong to the intermediate classes mostly between E and cD. There is a continuous distribution in the properties of the BCG's envelopes, ranging from undetected (E class) to clearly detected (cD class), with intermediate classes (E/cD and cD/E) showing the increasing degrees of the envelope presence. A minority ($\sim7$\%) of BCGs have disk morphologies, with spirals and S0s in similar proportions, and the rest ($\sim2$\%) are mergers. After carefully fitting the galaxies light distributions by using one-component (\sersic) and two-component (\sersic+Exponential) models, we find a clear link between the BCGs morphologies and their structures and conclude that a combination of the best-fit parameters derived from the fits can be used to separate cD galaxies from non-cD BCGs. In particular, cDs and non-cDs show very different distributions in the \re--\reff\ plane, where \re\ is the effective radius and \reff\ (the residual flux fraction) measures the proportion of the galaxy flux present in the residual images after subtracting the models. In general, cDs have larger \re\ and \reff\ values than ellipticals. Therefore we find, in a statistically robust way, a boundary separating cD and non-cD BCGs in this parameter space. BCGs with cD morphology can be selected with reasonably high completeness ($\sim 75\%$) and low contamination ($\sim 20\%$). This automatic and objective technique can be applied to any current or future BCG sample with good quality images. 

\end{abstract}


\section{Introduction} 
The brightest cluster galaxies (BCGs) are the most luminous and massive galaxies in today's universe. Their stellar masses reach beyond $\sim 10^{11} M_{\odot}$, and they reside at the bottom of the gravitational potential well of galaxy clusters and groups. Their formation and evolution relate closely to the evolution of the host clusters \citep{whiley08} and further tie to the history of large-scale structures in universe (\citealt{Conroy07}). BCGs are typically classified as elliptical galaxies \citep{LP92}, but a fraction of them possess an extended, low surface brightness envelope around the central region. These are referred to as cD galaxies (e.g. \citealt{Dressler84}; \citealt{OH01}). 

The surface brightness profile of elliptical galaxies was originally modelled using the empirical $R^{1/4}$ de Vaucouleurs law \citep{deVaucl48}. However, \cite{Lugger84} and \cite{Schombert86} showed that the $r^{1/4}$ model cannot properly describe the flux excess at large radii for most elliptical galaxies, and an additional parameter $n$ was introduced in the so-called \sersic\ ($r^{1/n}$) law \citep{Sersic63}. For the most massive early-type galaxies, however, a single \sersic\ profile still does not reproduce their luminosity distribution accurately. \citet{Gonzalez05} found that a sample of $30$ BCGs were best fitted using a double $r^{1/4}$ de Vaucouleurs profile rather than a single \sersic\ law. Furthermore, \citet{Donzelli11} suggested that a two-component model with an inner \sersic\ and an outer exponential profile is required to properly decompose the light distribution of $\sim48\%$ of the BCGs in their 430 galaxy sample. A similar conclusion was obtained by \citet{Seigar07}. 

The light profiles of BCGs need to be explained by any successful model of galaxy formation and evolution. In hierarchical  models of structure formation, a two-phase scenario is currently favoured. \cite{Hopkins09} proposed that a early central starburst could give rise to the bulge (elliptical) component of these galaxies, while the outer envelope was subsequently formed by the violent relaxation of stars originating in galaxies which merged with the central galaxy. Alternatively, \cite{Oser10} and \cite{Johansson12} suggested that intense dissipational processes such as cold accretion or gas-rich mergers could rapidly build up an initially compact progenitor and, after the star formation is quenched, a second phase of slower, more protracted evolution is dominated by non-dissipational processes such as dry minor mergers to form the low-surface-brightness outskirts. 

To shed light on the mechanism(s) leading to the formation of BCGs, especially of cD galaxies, we need to answer questions such as: are elliptical and cD BCGs two clearly distinct and separated classes of galaxies? if so, are elliptical and cD BCGs formed by different processes or in different environments? are the extended envelopes of cD galaxies intrinsically different structures which formed separately from the central bulge? To help answer these questions, in this paper we explore statistically how the visual classification of BCGs into different morphological classes (e.g., elliptical, cD; here referred to as ``morphology''), relates to the quantitative structural properties of their light profiles (e.g., effective radius $R_{\rm e}$, \sersic-index $n$; generically called ``structure'' in this paper). Moreover, finding an automatic and objective way to select cD BCGs is nontrivial for the future databases and study. Recent study such as \citet{Liu08} identified cD BCGs by Petrosian parameter profiles (\citealt{Petrosian76}), but their method does not give an unambiguous criterion to separate cD galaxies from non-cD BCGs. 

In this paper, we visually-classify $625$ BCGs from the sample of \citet[hereafter L07]{Linden07} and fit accurate models to their light profiles. We find clear links between the visual morphologies and the structural parameters of BCGs, and these allow us to develop a quantitative and objective method to separate cDs galaxies from ellipticals BCGs. In a later paper (Zhao et al., in preparation) we will study how the visual morphology and structural properties of BCGs correlate with their intrinsic properties (stellar masses) and their environment (cluster mass and galaxy density), and explore the implications that such correlations have for the formation mechanisms and histories of cDs/BCGs. 

The paper is organized as follows. In \S\ref{sec:data} we introduce the BCG samples and the visual morphological classification of the BCGs. In \S\ref{sec:BCGstruct} we describe the light distribution models and the fitting methods we use, and discuss how the results are affected by sky-subtraction uncertainties. This section also presents a quantitative evaluation of the quality of the fits. In \S\ref{sec:structureresult} we present the structural properties of the BCGs in the sample. In \S\ref{sec:border} we introduce an objective diagnostic to separate cDs from non-cD BCGs using quantitative information from their light profiles. We summarise our main conclusions in \S\ref{sec:conclude}.

\section{Data}
\label{sec:data}
\subsection{BCG Catalogue and Images}
To study the structural properties of BCGs in galaxy groups and clusters, we use the BCG catalogue published by L07. The groups and clusters that host these BCGs come from the C4 cluster catalogue \citep{Miller05} extracted from the Sloan Digital Sky Survey \citep[SDSS;][]{York00} third data release spectroscopic sample. The cluster-finding algorithm used to build the C4 catalogue identifies clusters as over-densities in a seven-dimensional parameter space of position, redshift and colour, minimising projection effects. The C4 catalogue gives a very clean widely-used cluster sample which is well supported by simulations. Based on the C4 catalogue, L07 restricted their cluster sample to the $0.02\leqslant z \leqslant 0.10$ redshift range to avoid problems related to the 55 arcsec ``fiber collision'' region of SDSS. Within each cluster, L07 applied an improved algorithm to identify the BCG as the galaxy being closest to the deepest point of the potential well of the cluster (see \citealt{Linden07} for a detailed discussion of this identification), and developed an iterative algorithm to measure the cluster velocity dispersion \vdvir\ within the virial radius $R_{200}$\footnote{$R_{200}$ is the radius within which the average mass density is $200\rho_{c}$, where $\rho_{c}$ is the critical density of the universe.}. The catalogue created by L07 contains $625$ BCGs in galaxy groups and clusters with redshifts $0.02\leqslant z \leqslant 0.10$ spanning a wide range of cluster velocity dispersions, from galaxy groups (\vdvir$\,\leqslant 200\,$\kms) to very massive clusters (\vdvir \,$\sim 1000\,$\kms).  $75\%$ of the BCGs in L07 are in dark matter halos with $\sigma_{r200} \geqslant 309\,$\kms, where the completeness of the halos identified by the C4 algorithm is expected to be above $50\%$. Obviously, for larger halo masses the completeness is higher. 

The images we use to classify the BCGs and analyse their structural properties come from the SDSS Seventh Data Release (DR7) $r$-band images. We also use SDSS-DR7 $g$-band images of the BCGs in Section~\ref{sec:uncertainty}. The BCG catalogue used in this paper together with their main properties are presented in appendix~A.

\subsection{Visual Classification}
\label{sec:visualclassification}
\begin{figure*}
\center{\includegraphics[scale=0.38, angle=0]{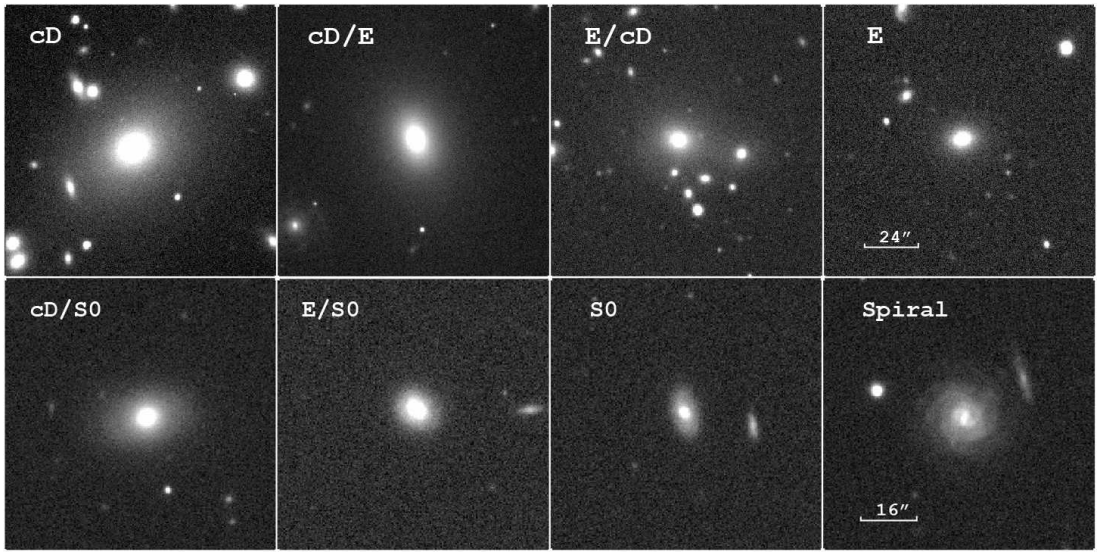}}
\caption{Examples of the main morphological classes of BCGs in our sample (cD, cD/E, E/cD, E, cD/S0, E/S0, S0, Spiral) illustrating the gradual transition between classes. The images are displayed using a logarithmic surface-brightness scale.}
\label{fig:atlas}
\end{figure*}

The 625 BCGs in L07 sample were visually classified by careful inspection of the SDSS images. BCGs were displayed using a logarithmic scale between the sky level and the peak of the surface brightness distribution. The contrast was adjusted manually to ensure that the low-surface-brightness envelopes were revealed if present. cD galaxies are identified by a visually extended envelope, while the envelope is not visible in our elliptical BCGs. Finally the BCGs were classified into three main types: 414 cDs, including pure cD (356), cD/E (53) and cD/S0 (5); 155 ellipticals, including pure E (80), E/cD (72), and E/S0 (3); 46 disk galaxies, containing spirals (24) and S0s (22). The main morphological classes of BCGs are illustrated in Figure~\ref{fig:atlas}. There are also 10 BCGs undergoing major mergers, but we will not discuss them in this paper in any detail. 

Over half of the BCGs in the sample are classified as cDs. Separating cD BCGs and non-cD elliptical BCGs is a very hard problem since there is no sharp distinction between these two classes \citep[e.g.,][]{Patel06,Liu08}. Detecting the extended stellar envelope that characterises cD galaxies depends not only on its dominance, but also on the quality and depth of the images, and on the details of the method(s) employed. We used intermediate classes such as cD/E (probably a cD, but could be E) and E/cD (probably E, but could be cD) to account for the uncertainty inherent in the visual classification.

Our careful inspection of the images clearly reveals that there is a wide range in the brightness and extent of the envelopes. There seems to be a continuous distribution in the envelope properties, ranging from undetected (pure E class) to clearly detected (pure cD class), with the intermediate classes (E/cD and cD/E) showing increasing degrees of envelope presence. This continuous distribution in envelope detectability will also be made evident in the structural analysis carried out later in this paper. The classification we present here does not intend to be a definitive one since such a thing is probably unachievable. Our aim is to obtain a homogeneous and systematic visual classification of the BCGs and then study how such classification correlates with quantitative and objective structural properties of the BCGs. The visual morphological types of all the galaxies in the sample are presented in Appendix~A.  

\begin{figure}
\centering{\includegraphics[scale=0.7]{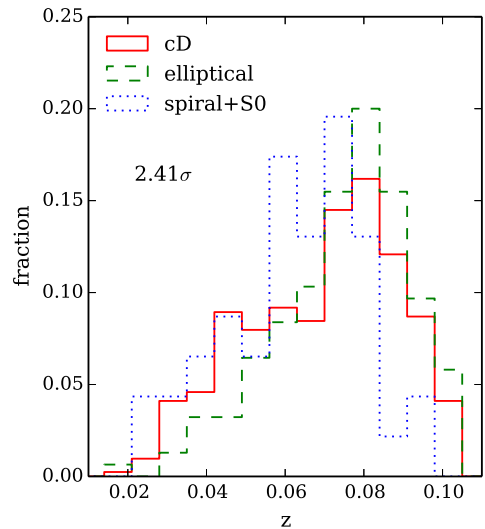}}
\caption{Redshift distribution for BCGs with different morphological types. The red solid line corresponds to cD BCGs, the green dashed line to ellipticals, and the blue dotted line to disk (spiral and S0) BCGs. A Kolmogorov-Smirnov test
indicates that the redshift distributions of cD and elliptical BCGs are only different at the $\sim2.4\sigma$ level. cD galaxies share the same redshift distribution with elliptical BCGs at $z\geqslant0.05$, while there are proportionally $\sim 10\%$ more cD galaxies at $z<0.05$.}
\label{fig:redshift}
\end{figure}

We checked the effect that the redshift of BCGs may have on the visual classification. cDs might be mistakenly identified as elliptical if they are more distant since the extended low-surface-brightness envelope may be harder to resolve at higher redshifts. Figure~\ref{fig:redshift} illustrates the redshift distribution of the three main types. cD galaxies generally share the same redshift distribution with elliptical BCGs, especially at $z\geqslant0.05$. At $z<0.05$ we identify a slightly higher proportion (by $\sim10$\%) of cD galaxies. However, if we compare the structural properties of cD and elliptical BCGs which are at $z\geqslant0.05$, the results we obtain do not significantly differ from those using the full-redshift sample. As an additional check, we artificially redshifted some of the lowest redshift galaxies ($z\sim0.02$--$0.03$) to $z=0.1$, the highest redshift of the sample, taking into account cosmological effects such as surface-brightness dimming. Because the redshift range of the BCGs we study is very narrow, the effect on the images is minimal and does not have any significant impact on the visual classification. We are therefore confident that our visual classification is robust and that in the relatively narrow redshift range explored here any putative redshift-related biases will not affect our results.

\section{Quantitative Characterisation of BCG Structure}
\label{sec:BCGstruct}

The surface brightness profiles of galaxies provide valuable information on their structure and clues to their formation. It has become customary to fit the radial surface brightness distribution using theoretical functions which have parameters that include a measurement of size (e.g., half-light radius or scale length), a characteristic surface brightness, and other parameter(s) describing the shape and properties of the surface brightness profiles. In this paper we use GALFIT (\citealt{Peng02}) to fit the 2-D luminosity profile of each BCG using two parametric models, and thus determine the best-fitted parameters of each model. GALFIT can simultaneously fit model profiles to several galaxies in one image, which is particularly important for BCGs since they usually inhabit very dense environments. In this way, the light contamination from nearby galaxies can be accounted for appropriately.

We explore two models to represent the luminosity profile of the BCGs. A model commonly used to fit a variety of galaxy light profiles is the generalization of the $r^{1/4}$ \cite{deVaucl48} law introduced by \cite{Sersic63}. The \sersic\ model has the form 
\begin{equation}
\label{eq:sersic}
I(r)=I_{\rm e} \exp \{-b[(r/r_{\rm e})^{1/n}-1]\},
\end{equation}
where $I(r)$ is the intensity at distance $r$ from the centre, \re, the effective radius, is the radius that encloses half of the total luminosity, $I_e$ is the intensity at \re, $n$ is the \sersic \ index representing concentration, and $b \simeq 2n-0.33$ (\citealt{Caon93}). The \sersic\ function provides a good model for galaxy bulges and massive elliptical galaxies. Since BCGs are mostly early-type galaxies, it is reasonable to fit their structure with single \sersic\ models first. Subsequently, in order to explore the complexity introduced by the extended envelopes of cD galaxies, we will also fit the light profile of BCGs adding an additional exponential component to the \sersic\ profile. Adding this exponential component is the simplest way to describe the ``extra-light'' from the extended envelope. Note that the exponential profile $I(r)=I_0 \exp (-r/r_{\rm s})$ is just a \sersic\ model with $n=1$. The models assume that the isophotes have elliptical shapes, and the ellipticity and orientation of each model component are parameters determined in the fitting process. 

In order to run GALFIT, we require a postage stamp image for each BCG with appropriate size to measure its structure over the full extent of the object, a mask image with the same size as the stamp image, an initial guess for the fitting parameters, an estimate of the background sky level, and a point spread function (PSF). Details on how these ingredients are produced and the fitting procedures are given below.

\subsection{Pipeline for One-Component Fits: Modified GALAPAGOS}
\label{sec:galapagos}

We run GALFIT using the GALAPAGOS pipeline \citep{Barden12}. GALAPAGOS has been successfully applied to a wide variety of ground- and space-based images \citep{Haubler13,Vika13,Vulcani14}. \citet{Guo09} specifically tested a modified version of GALAPAGOS to fit the central galaxies of local clusters using SDSS $r$-band images. We applied the same modified version of GALAPAGOS to fit the SDSS $r$-band images of the BCGs in our sample. The starting point are SDSS images with a size of $2047 \times 1488$ pixels. For each BCG, the pipeline carries out four main tasks before running GALFIT itself: (i) detection of all the sources present in the image; (ii) cutting out the appropriate postage stamp and preparing the mask image; (iii) estimation of the sky background; (iv) preparation of the input file for GALFIT. After completing these tasks, GALAPAGOS will run GALFIT using the appropriate images and input parameters. We describe now these tasks in detail.

\textit{(i) Source Detection:} SExtractor (\citealt{BA96}) is used to detect galaxies in the SDSS images. A set of configuration parameters defines how SExtractor detects sources. The values of the SExtractor input parameters follows \citet{Guo09}: DETECT\_MINAREA$\,=25$, DETECT\_THRESH$\,=3.0$, and DEBLEND\_MINCONT$\,=0.003$. This set of parameters were tested to perform well on SDSS $r$-band images so that the bright and extended BCGs were isolated from other sources without artificially deblending them into multiple components. SExtractor also provides  estimates of several properties for the target BCGs and nearby objects such as their magnitude, size, axis ratio and position angle. These values are used to calculate the initial guesses of the model parameters that are needed as inputs by GALFIT.

\textit{(ii) Postage Stamp creation:} GALAPAGOS cuts out a rectangular postage stamp centred on the target BCG which will be used by GALFIT as input image.  We define the ``Kron ellipse'' for a galaxy image as an ellipse whose semi-major axis is the Kron radius\footnote{In this paper we use the following definition of ``Kron radius'': \rkron$\,=2.5r_1$, where $r_1$ is the first moment of the light distribution \citep{Kron80,BA96}. For an elliptical light distribution, this is, strictly speaking, the semi-major axis.} (\rkron), with the ellipticity and orientation determined by SExtractor. The postage stamp size is determined in such a way that it will fully contain an ellipse 3.5 times larger than the Kron ellipse, i.e., its semi-major axis is $3.5$\rkron, and has the same ellipticity and orientation. The 3.5 factor represents a compromise between computational speed and ensuring that virtually all the BCG's light is included in the postage stamp. At this stage, a mask image is also created, identifying and masking out all pixels belonging to objects in the postage stamp which will not be simultaneously fitted by GALFIT. The aim is to reduce the computational time by excluding objects too far from the BCG or too faint to have any significant effect on the fit. Following \cite{Barden12}, an ``exclusion ellipse'' is defined for each galaxy with a semi-major axis $1.5$\rkron$\, + 20\,{\rm pixels}$, and the same ellipticity and orientation as the Kron ellipse. GALAPAGOS masks out all objects whose exclusion ellipse does not overlap with the exclusion ellipse of the target BCG. These objects are deemed to be too far away from the BCG to require simultaneous fitting. Furthermore, all objects more than $2.5\,$magnitudes fainter than the BCG are also masked out since they are too faint to affect the BCG fit. The pixels that belong to these objects according to the SExtractor segmentation maps are masked out and excluded from the fits. All the remaining objects will be simultaneously fitted by GALFIT at the same time as the BCG. For a detailed description of this process and a justification of the parameter choice see \cite{Barden12}.

\textit{(iii) Sky Estimation:} Accurate estimates of the sky background level is crucial when fitting galaxy profiles, particularly when interested in the low-surface-brightness outer regions. Overestimating the sky level will result in the underestimation of the galaxy flux, size, and \sersic\ index $n$, and vice-versa. GALAPAGOS uses a flux growth curve method to robustly estimate the local sky background around the target galaxy. SDSS DR7 also provides a global sky value for the whole $2047 \times 1488$ image frame and local sky values for each galaxy. The SDSS \textit{PHOTO} pipeline estimates the sky background using the median flux of all the pixels in the image after $2.33\sigma$-clipping. However, according to the SDSS-III website, the version of \textit{PHOTO} used in DR7 and earlier data releases tended to overestimate both the global and local sky values. The sky measurement is improved by SDSS-III in later data releases, but since we use the images from DR7 we cannot use the SDSS sky value with enough confidence. \citet{Haubler07} demonstrated that the sky measurement that GALAPAGOS produces is highly reliable for single-band fits because it takes into account the effect of all the objects in the image. Therefore, in this study we use the local sky background estimated by GALAPAGOS. The accurate sky measurement provided by GALAPAGOS indicates that we can reach a surface brightness limit in the $r$-band of $\sim 27$ mag/arcsec$^{2}$. This is deep enough to study the faint extended structures of BCGs. For each galaxy, its local sky background is included in the GALFIT input file and is fixed during the fitting procedure. Given the importance of accurate sky subtraction, in Section~\ref{sec:skyuncer} we will carry out an explicit comparison of our results using SDSS and GALAPAGOS sky estimates.

\textit{(iv) GALFIT Input:} GALAPAGOS produces an input file which includes initial guesses for the fitting parameters based on the SExtractor output. As mentioned above, all objects which are not masked out are fitted simultaneously using a \sersic\ model. The initial-guess model parameters for these nearby companions are also determined from SExtractor. In order to obtain reasonable results, we impose some constraints on the acceptable model parameter range.  Our constrains on position, magnitude, axis ratio and position angle follow \citet{Haubler07}. Additionally, the half-light radius \re\ is constrained within $0.3 \leqslant \rex \leqslant 800\,$pixels. This prevents the code from yielding unreasonably small or large sizes. Since the pixel size of the SDSS images is $0.396\,$arcsec, \re\ is constrained to be larger than $0.12\,$arcsec, which is much smaller than the PSF, and smaller than half the size of the original input images, reasonable for the range of redshifts explored. In the original GALAPAGOS pipeline, the constraint on the \sersic\ index is $0.2 \leqslant  n \leqslant 8$. These are reasonably conservative limits, since normal galaxies with $n>8$ are rarely seen and are often associated with poor model fits. However, some studies have shown that very luminous elliptical galaxies with $n>8$ do exist \citep[e.g.,][]{Graham05}, therefore for the target BCGs we allow $n$ to be as large as 14 to keep the fits as free as possible. For the companion galaxies, which are fitted simultaneously, we still keep the constraint $0.2 \leqslant  n \leqslant 8$. The final ingredient needed by GALFIT is a PSF image appropriate for each BCG. These are extracted from the SDSS DR7 data products\footnote{http://www.sdss.org/DR7/products/images/read\_psf.html} according to the photometric band used and the position of the BCG on the SDSS image.

\subsection{Effect of the Sky Background Subtraction: Comparing  SDSS and GALAPAGOS Sky Estimates}
\label{sec:skyuncer}

\begin{figure}
\raggedright{\includegraphics[scale=0.75]{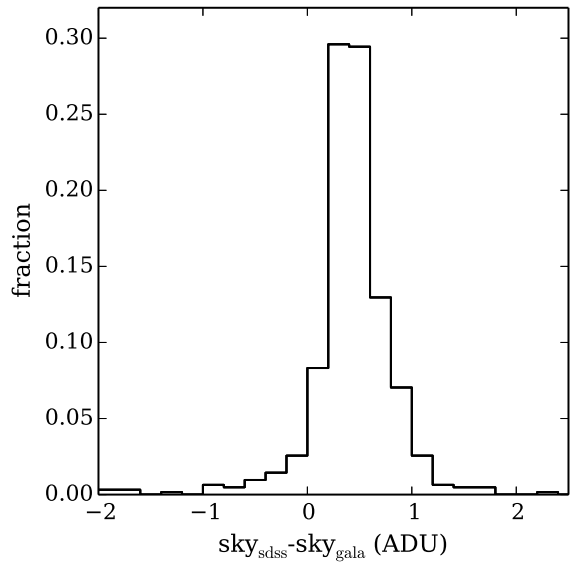}}
\caption{Distribution of the difference between the SDSS DR7 global sky and the GALAPAGOS-measured sky values. In general, SDSS overestimates the sky background. The average sky value measured by GALAPAGOS in the SDSS $r$-band BCG images is $140.8$ ADU, corresponding to a surface brightness of $\sim 20.9$ mag/arcsec$^{2}$.} 
\label{fig:skydiff}
\end{figure}

\begin{figure}
\raggedright{\includegraphics[scale=0.56]{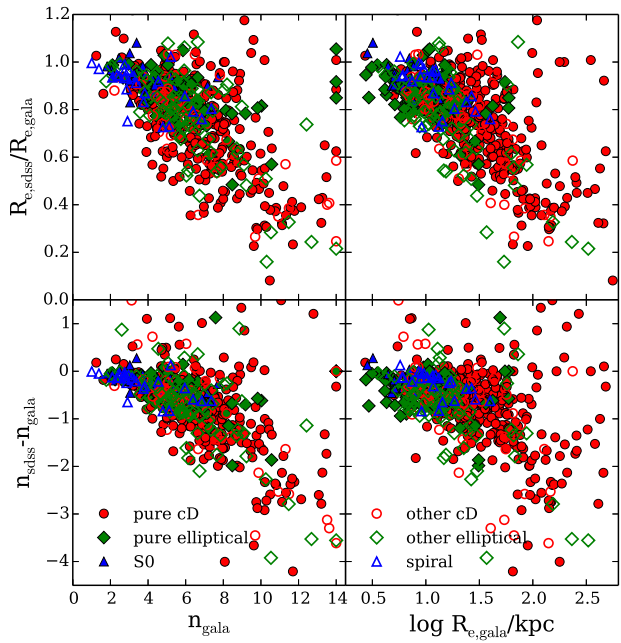}}
\caption{Comparison on the best-fit $n$ and \re\ from single \sersic\ models using the SDSS and GALAPAGOS-measured sky estimates. Solid and open red circles correspond to pure cD and other cD galaxies (cD/E and cD/S0) respectively; solid and open green diamonds correspond to pure and other (E/cD and E/S0) elliptical BCGs respectively; solid blue triangles represent S0s and open ones are spirals. It shows that the SDSS overestimation of the global sky result in the values of $n_{\rm sdss}$ and $r_{\rm e,sdss}$ being smaller than the corresponding GALAPAGOS ones. Moreover, the effect is more serious for the BCGs with large $n$ and \re\ which are mostly cDs.}
\label{fig:skyparam}
\end{figure}

As described in Section~\ref{sec:galapagos}, in this study we rely on the sky measurements provided by GALAPAGOS. However, it is important to test the effect that the choice of sky background has on our results. We do this by comparing the fitted \sersic\ model parameters $n$ and \re\  using the GALAPAGOS and SDSS sky estimates. As mentioned before, SDSS DR7 provides a global sky value for the whole $2047 \times 1488$ image and local sky values for each galaxy. \citet{Guo09} found that the local background estimates are generally larger than the global ones due to contamination from the outskirts of extended and bright sources, making them unreliable. We therefore restrict our comparison to the global SDSS sky values. We fit the BCG light profiles twice using exactly the same procedure and input parameters (see \S\ref{sec:galapagos}) but changing only the sky background estimates. The first set of fits use the GALAPAGOS-determined sky values, while the second set use the SDSS DR7 global ones.

Figure~\ref{fig:skydiff} shows the distribution of the difference between the SDSS DR7 global sky and the sky measured by GALAPAGOS. It is clear that the SDSS global sky is generally larger than the local sky from GALAPAGOS. The effect from different sky values on the best-fitted structural parameters (\sersic\ index $n$ and effective radius \re) is shown in Figure~\ref{fig:skyparam}.  It is clear that the SDSS larger sky values result in the values of $n_{\rm sdss}$ and $r_{\rm e,sdss}$ being smaller than the corresponding GALAPAGOS ones. The effect becomes more severe for those BCGs with large $n$ and \re, most of which are cD galaxies. This means the overestimated sky values would particularly affect the measurements on the low-surface-brightness envelopes of cD galaxies. Although it is difficult to know \textit{a priori} which the \textit{true} value of the sky background is, based on the fact that the SDSS-III provides evidence that DR7 sky values are overestimated while \citet{Haubler07} showed reasonable proof of the reliability of the GALAPAGOS sky measurements, in what follows we will therefore trust and use the GALAPAGOS-determined sky values.

\subsection{Two-Component Fits}
\label{sec:BDfit}
Although the light profiles of many early-type galaxies can be reproduces reasonably well with single \sersic\ models, the extended envelopes of cD galaxies may require an additional component. We therefore fitted all the BCGs using a two-component model consisting of a \sersic\ profile plus an exponential. The input postage stamp, mask image, PSF, and sky values required by GALFIT remain the same as for the single-\sersic\ fits. To ensure that we are fitting exactly the same light distribution, the location of the centre of the BCG is fixed to the X and Y coordinates determined in the single fit, and we also force the initial guesses of the model parameters to be the single-component fit results. The BCG companions are simultaneously fitted still with single-\sersic\ profiles but with initial-guess parameters determined by the single profile fits. 

\subsection{Residual Flux Fraction and Reduced $\chi^2$}
\label{sec:rff}
Although the models we are fitting are generally reasonably good descriptions of the BCG light profiles, real galaxies can be more complicated, with additional features and structures such as star-forming regions, spiral arms, and extended halos. It is therefore desirable to quantify how good the fits are and what residuals remain after subtracting the best-fit models. A visual inspection of the residual images can generally give a good feel for how good a fit is, and sometimes tell us whether an additional component or components are required. However, more quantitative, repeatable and objective diagnostics are also needed. The residual flux fraction \citep[\reff;][]{Hoyos11} provides one such diagnostic. It is defined as 
\begin{equation}
\label{eq:rff}
RFF=\dfrac{\sum_{i,j\in A} |I_{i,j}-I_{i,j}^{\rm model}|-0.8\times \Sigma_{i,j\in A}\sigma_{i,j}^{\rm bkg}}{\Sigma_{i,j\in A} I_{i,j}},
\end{equation}
where $A$ is the particular aperture used to calculate \reff. Within A, $I_{i,j}$ is the original flux of pixel $(i.j)$, $I_{i,j}^{\rm model}$ is the model flux created by GALFIT, and $\sigma_{i,j}^{\rm bkg}$ is the \textit{rms} of the background. \reff\ measures the fraction of the signal contained in the residual image that cannot be explained by background noise. The $0.8$ factor ensures that the expectation value of the \reff\ for a purely Gaussian noise error image of constant variance is $0.0$. See \cite{Hoyos11} for details. Obviously, this diagnostic can be applied to both single-\sersic\ and two-component profiles, or any other model. The aperture A we use to calculate \reff\ is the ``Kron ellipse'' defined in Section~\ref{sec:galapagos} (an ellipse with semi-major axis \rkron\ and the ellipticity and orientation determined by SExtractor for the BCG). $\Sigma_{i,j\in A} I_{i,j}$, the denominator of Equation~(\ref{eq:rff}), is computed as the total BCG flux contained inside the Kron ellipse, which is one of the SExtractor outputs, and therefore independent of the model fit. 

Since BCGs usually reside in dense environments, sometimes there are some faint nearby objects contained within the Kron ellipse that have not been fitted by GALFIT (those more than $2.5\,$mag fainter than the BCG, see \S\ref{sec:galapagos}). These objects will be present in the residual image.  Moreover, brighter companions that have been simultaneously fitted may also leave some residuals due to inaccuracies in their fits. Therefore, even if the BCG light distribution has been accurately fitted, \reff\ can be affected by the residuals from the companion galaxies, failing to provide an accurate measure of the quality of the fit. To minimise the effect from companion galaxies on \reff, we mask out the pixels belonging to all companions within the Kron ellipse using SExtractor segmentation maps. The \reff\ will therefore measure the residuals from the BCG fit alone, excluding, as far as possible, those belonging to nearby galaxies.

An additional measurement of the fit accuracy is the reduced $\chi^2$, which is minimised by GALFIT when finding the best-fit models. It is defined as 
\begin{equation}
\label{eq:chi2}
\chi^2_\nu=\frac{1}{N_{\rm dof}} \sum_{i,j\in A} \frac{(I_{i.j}-I^{\rm model}_{i,j})^2}{\sigma^2_{i,j}},
\end{equation}
where $A$ is the aperture used to calculate $\chi^2_\nu$, $N_{\rm dof}$ is the number of degrees of freedom in the fit, $I_{i,j}$ is the original image flux of pixel $(i,j)$. $I^{\rm model}_{i,j}$ represents, for each pixel, the sum of the flux of the models fitted to all the galaxies in the aperture, and  $\sigma_{i,j}$ is the noise corresponding to pixel $(i,j)$. This noise is calculated by GALFIT taking into account the contribution of the Poisson errors and the read-out-noise of the image \citep{Peng02}. 

Similarly to \reff, \chisq\ also measures the deviation of the fitted model from the original light distribution. The value of \chisq\ that GALFIT minimises to find the best-fit model is calculated over the whole postage stamp, and includes contributions from all the objects fitted. To make sure that we only take into account the contribution to \chisq\ from the BCG fit, we calculate it within the Kron ellipse of the BCG, masking out the nearby objects as we did when calculating \reff.

The choice of aperture (Kron ellipse with semi-major axis of \rkron) over which we evaluate \reff\ and \chisq\ represents a good compromise between covering a large fraction of the galaxy light while minimising the impact of close companions. We carried out several tests to evaluate the sensitivity of our results to the changes in aperture size. If we reduce the semimajor axis of the aperture by $20$\% or more we lose significant information on the extended halo of BCGs, which we must avoid. If we increase the semimajor axis of the aperture by $20$\% or more, we potentially increase the sensitivity to the galaxy halos but in the crowded central cluster regions contamination from companion galaxies becomes a serious problem, generally increasing \reff\ and \chisq. Changes in the aperture semimajor axis within $\pm20$\% would have no effect on the conclusions of this paper.

\subsection{Evaluating One-Component and Two-Component Fits}
\label{sec:1cvs2c}

\begin{figure*}
\center{\includegraphics[scale=0.4, angle=0]{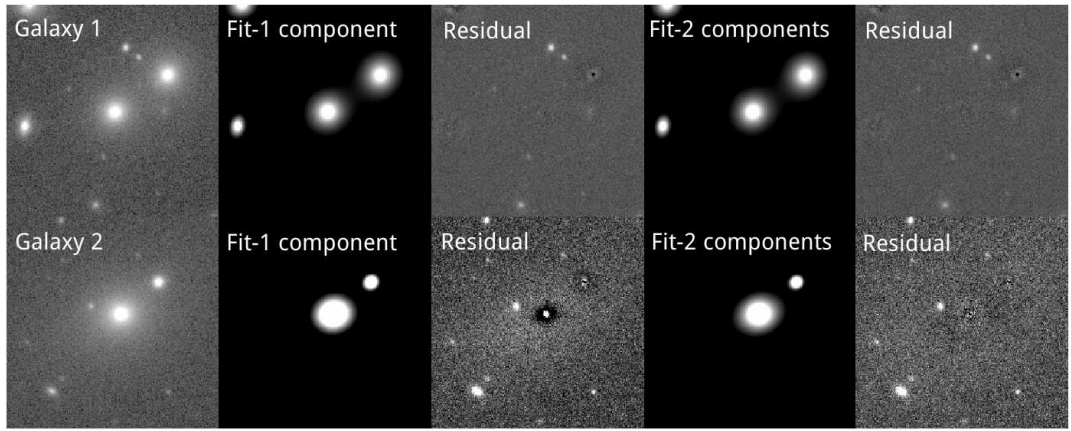}}
\caption{Example of one-component (\sersic) fits and two-component (\sersic+Exponential) fits for 1C and 2C BCGs, respectively. From left to right, the panels show the original image, the one-component model, the residuals after subtracting the one-component fit, the two-component model, and the residuals after subtracting the two-component fit. The upper panels show a 1C BCG where a one-component fit does a good job and adding a second component does not visibly improve the residuals. The lower panels show a 2C BCG, where the one-component residual exhibits clear excess light at large radii, suggesting that a second component is necessary. Indeed, the two-component residual is much better for this BCG. } 
\label{fig:exp12BCG}
\end{figure*}

Since \reff\ and \chisq\ can quantify the residual images after subtracting the model fits, we attempt to use them to assess whether a one-component (\sersic) fit or a two-component (\sersic+Exponential) fit is more appropriate to describe the light profile of individual BCGs. In order to do this, we first evaluate the effectiveness of \reff\ and \chisq\ at quantifying the goodness-of-fit. We visually examine the fits and residuals obtained from both one- and two-component models for all the BCGs in our sample. In some cases, two of which are illustrated in Figure~\ref{fig:exp12BCG}, it is obvious which model is clearly favoured. 

For those BCGs where such a clear distinction can confidently be made, we classify them into what we call 1C (one-component) BCGs and 2C (two-component) BCGs. Explicitly, 1C BCGs (e.g., galaxy 1 in the top panel of Figure~\ref{fig:exp12BCG}) are those for which a one-component \sersic\ model represents their light distribution very well, and therefore the residuals left are small and show no significant visible structure. For these galaxies, adding a second component does not visibly improve the residuals. Conversely, 2C BCGs (e.g., galaxy 2 in the bottom panel of Figure~\ref{fig:exp12BCG}) are not well fitted by a one-component model, and the residuals are significant. These residuals often show excess light at large radii which can be identified as an exponential component or halo. Additionally, the fit to these galaxies visibly improves when using a two-component model. With these criteria we confidently identify $53$ 1C BCGs and $25$ 2C BCGs. Since we want to test the sensitivity of \reff\ and \chisq, we concentrate for now on this small but robust subsample. The rest of the BCGs (537) cannot be confidently classified into 1C or 2C BCGs because it is too hard to tell visually due to the residuals containing significant structures which cannot be accurately fitted by such simple models.

\begin{figure}
\subfigure{
\raggedright{\includegraphics[scale=0.25]{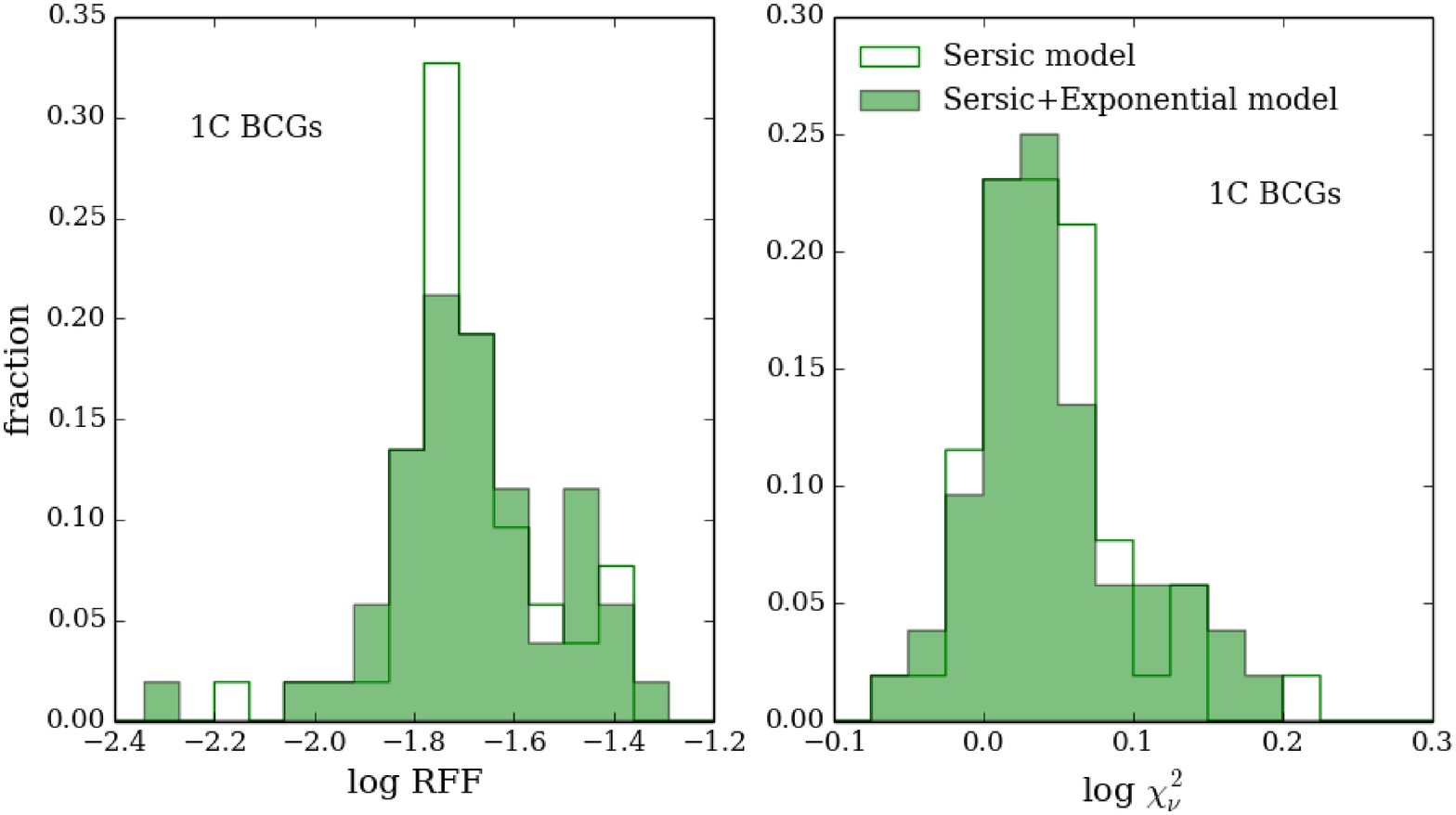}}}  
\subfigure{
\raggedright{\includegraphics[scale=0.25]{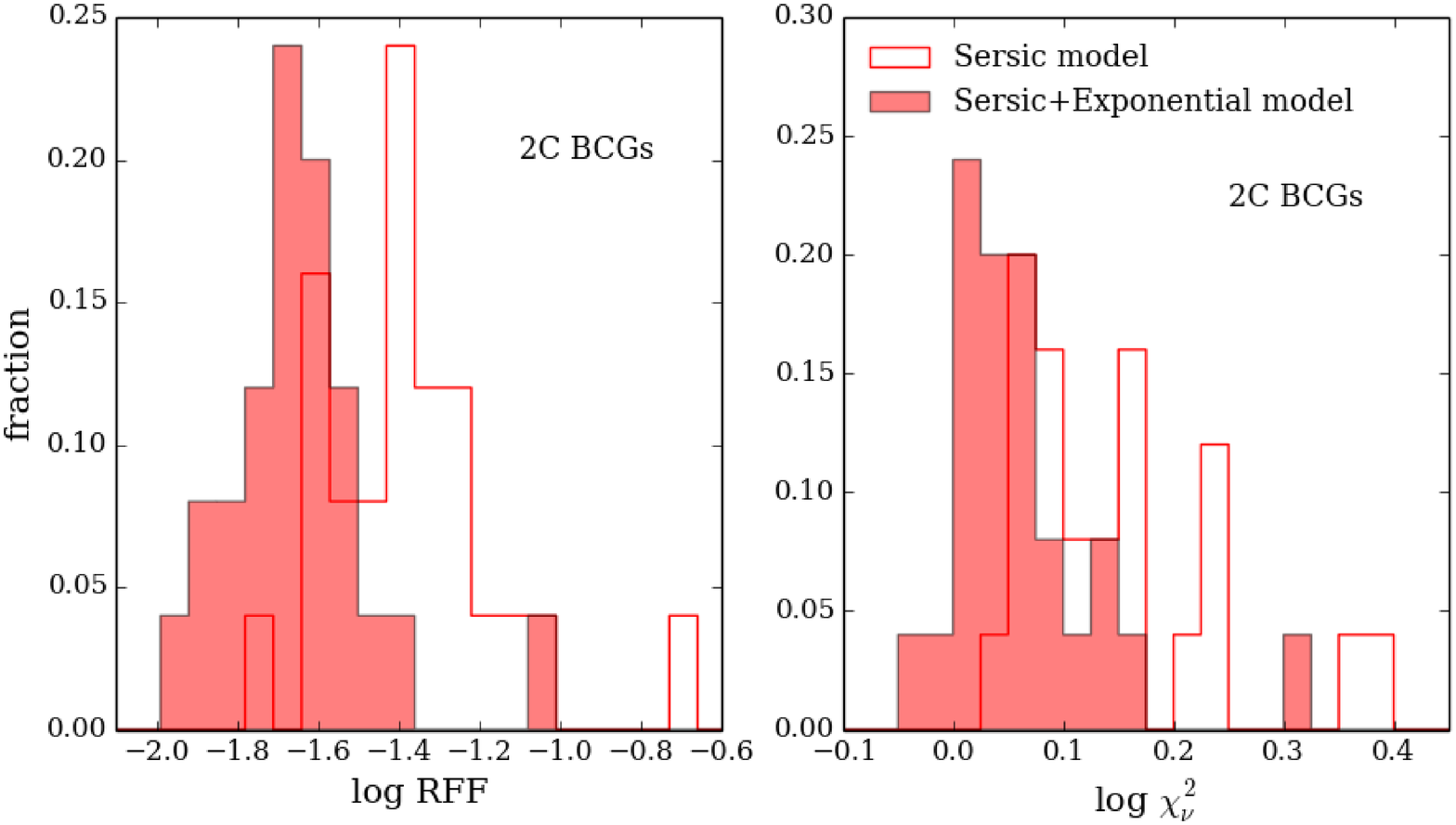}}}
\subfigure{
\raggedright{\includegraphics[scale=0.348]{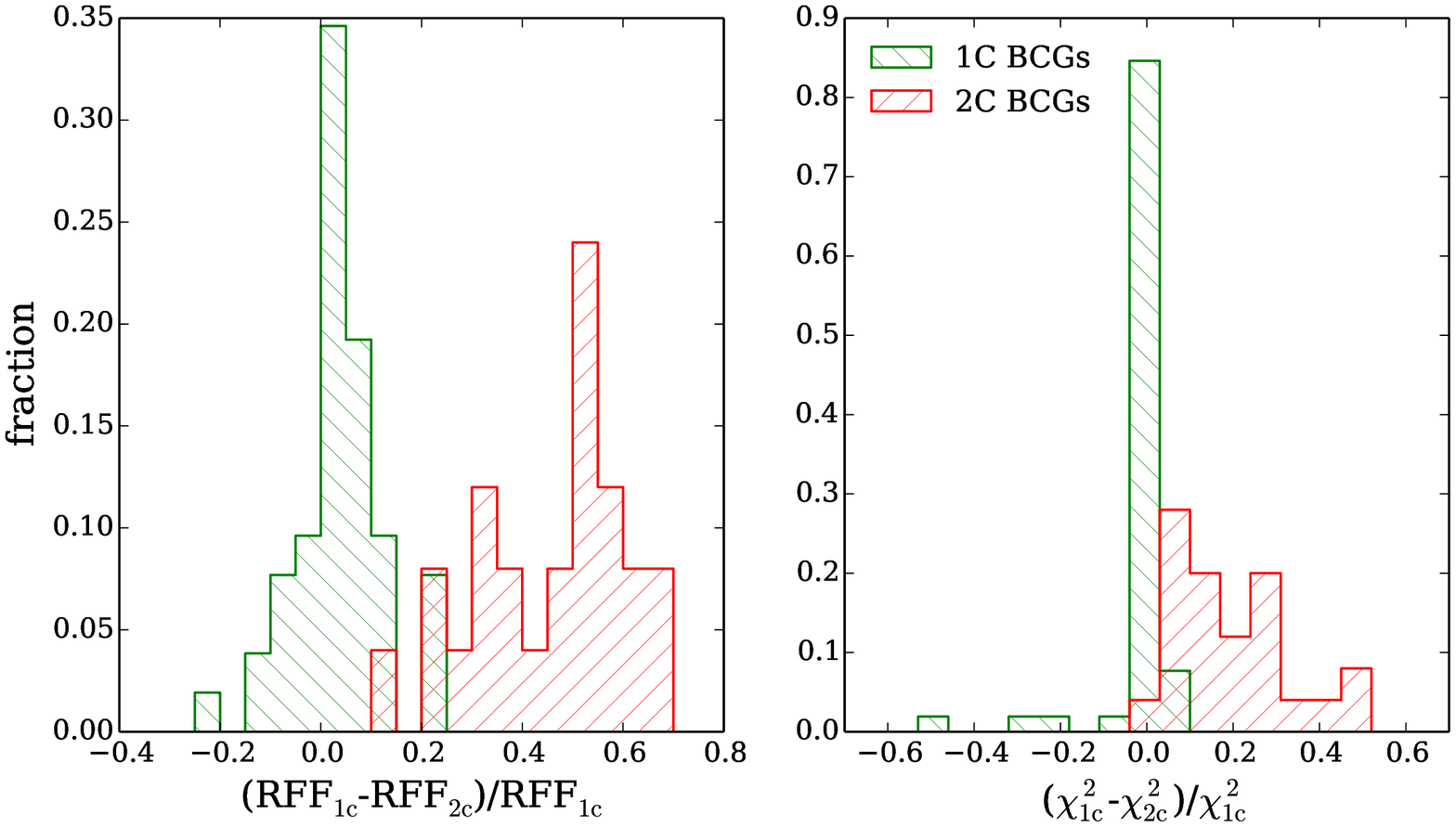}}}
\caption{The top four panels show the distribution of $\log$\reff\ (left) and $\log$\chisq\ (right) for single \sersic\ (open histograms) and \sersic+Exponential (solid histogram) fits. The two uppermost panels correspond to the $53$ 1C BCGs, while the middle panels correspond to the $25$ 2C BCGs. The two bottom panels show the difference in \reff\ and \chisq\ between one-component and two-component models for both sets of BCGs. Clearly, the \reff\ and \chisq\ distributions of one- and two-component fits are virtually indistinguishable for 1C BCGs. However, \reff\ and \chisq\ tend to be significantly smaller for the two-component fits of 2C BCGs. Typical values for good fits are $\log$\reff$\,\simeq -1.7^{+0.11}_{-0.06}$, and $\log$\chisq$\,\simeq 0.042^{+0.033}_{-0.025}$ (median $+$/$-$ the 1st and 3rd quartiles of the parameters). Both \reff\ and \chisq\ are sensitive to the magnitude of the residuals, but \reff\ is appears to be significantly more sensitive.}
\label{fig:comp12BCG}
\end{figure}

Figure~\ref{fig:comp12BCG} presents a comparison of the \reff\ and \chisq\ values for the one- and two-component fits of the $53$ 1C BCGs and $25$ 2C BCGs. For 1C BCGs, the \reff\ and \chisq\ distributions of one- and two-component fits are virtually indistinguishable. Neither \reff\ nor \chisq\ improve significantly when the second component is added. However, \reff\ and \chisq\ are significantly smaller for the two-component fits of 2C BCGs. It is clear therefore that the quantitative information that \reff\ and \chisq\ provide agrees very well with the visual assessments of the fits. Both \reff\ and \chisq\ are sensitive to changes in the residuals, but \reff\ appears to be more sensitive. As shown in the bottom panels of Figure~\ref{fig:comp12BCG}, the improvement in the two-component fit for 2C BCGs is around $40\%$--$60\%$ when measured by \reff, while it is only $\sim 20\%$ when measured by \chisq . A further useful piece of information obtained from this test is that the typical values of $\log$\reff\ and $\log$\chisq\ for fits deemed to be good by visual inspection are $\log$\reff$\,\simeq -1.7^{+0.11}_{-0.06}$, and of $\log$\chisq$\,\simeq 0.042^{+0.033}_{-0.025}$ (median $+$/$-$ 1st and 3rd quartiles of the parameter distributions).

As mentioned before, the majority of the BCGs cannot be visually classified into 1C or 2C BCGs with high certainty because their light distributions are too complex to be accurately represented by such simple models. Nevertheless, we can use the quantitative information provided by \reff\ and \chisq\ to gauge to what extent the BCGs are better fit by a two-component model than by a one-component model. This will be discussed later. 

We would like to point out that this is the first time that the residual flux is calculated considering \textit{only} the contribution of the target galaxies when estimating both \reff\ and \chisq, explicitly excluding the contribution due to the companion galaxies. For instance, \citet{Hoyos11} also used \reff\ to evaluate the goodness-of-fit, but they measured the residuals over all pixels within a specific area around the target galaxies, without excluding nearby companions. Similarly, the \chisq\ values from GALFIT have also been applied to evaluate which fitting model is better (e.g., \citealt{Bruce12}), but the effect of nearby objects on the \chisq\ values was also overlooked.  Using the 2C BCG sample, we assessed the importance of this improvement. If the \reff\ and \chisq\ are calculated considering the residuals in all the pixels inside the relevant aperture, the \reff\ and \chisq\ distributions for the two-component fits of 2C BCGs cannot be distinguished from the one-component results. The effect of the contribution to the residuals from companion galaxies is so severe that it renders such a comparison useless. Our method therefore represents a significant step forward. It is extremely important to exclude the contibution of the companion galaxies when calculating \reff\ and \chisq\ in this kind of analysis.

\section{Structural Properties of BCGs}
\label{sec:structureresult}

Our morphologically-classified BCGs provide a large sample to statistically study their structural properties and link them to their morphological properties. In what follows we consider the three main morphological classes of BCGs: cDs (including all BCGs classified as pure cD, cD/E and cD/S0); ellipticals (including pure E, E/cD and E/S0) and disk (spiral and S0) BCGs. The $10$ BCGs classified as mergers are excluded (see Section~\ref{sec:visualclassification} for details). We decided to include the galaxies with ``uncertain'' morphologies (such as cD/E and E/cD) in our analysis to reflect the difficulties involved in visual classification. However, to ensure the robustness of our analysis, at every stage we have checked that considering only ``pure'' cD and elliptical BCGs (i.e., excluding the cD/E, cD/S0, E/cD and E/S0 classes) would not change our conclusions.

Since most BCGs are early-type galaxies, we will first consider and discuss single \sersic\ models when fitting their SDSS $r$-band images. We will subsequently use \sersic+Exponential models to see whether the fits are improved. But before embarking in the analysis of the parameters derived from these model fits, we first evaluate their uncertainties.

\subsection{Structural Parameter Uncertainties}
\label{sec:uncertainty}

\begin{figure*}
\centering{\includegraphics[scale=0.65]{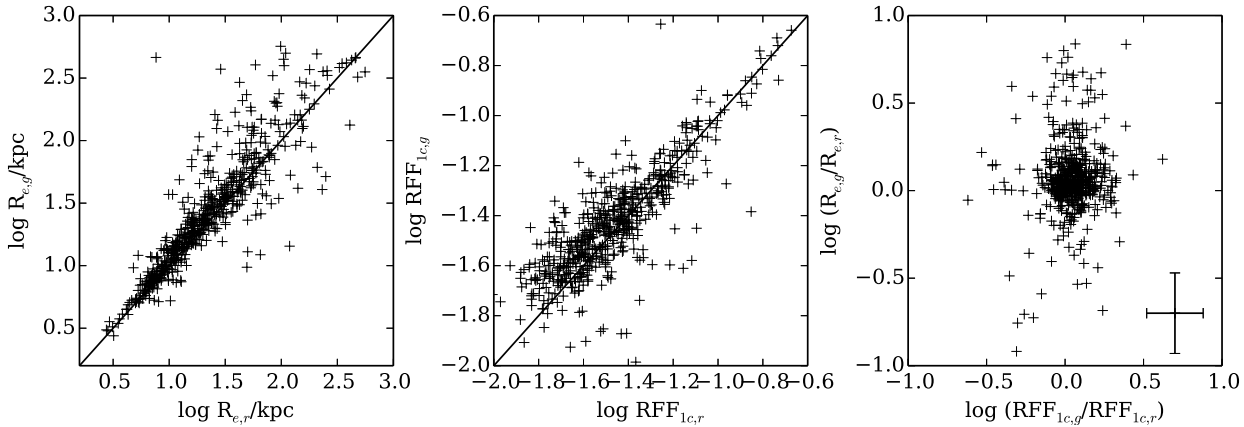}}
\caption{Comparison of the \re\ (left panel) and \rffs\ (middle panel) values obtained in both the SDSS $g$-band and $r$-bands. The solid lines correspond to the 1-to-1 relations. The right panel shows $\log (R_{\rm e,g}/R_{\rm e,r})$ vs.\ $\log (RFF_{\rm 1c,g}/RFF_{\rm 1c,r})$. The error bars in the bottom-right corner are derived from the \textit{rms} scatter of each parameter.}
\label{fig:error}
\end{figure*}

The parameter uncertainties that GALFIT reports are calculated using the covariance matrix derived from the Hessian matrix computed by the Levenberg-Marquardt algorithm that the program uses \citep{Peng10}. These formal uncertainties are only meaningful when the model provides a good fit to the image, in which case the fluctuations in the residual image are only due to Poisson noise. However, for real galaxy images the residual images contain not only Poissonian noise, but also systematics from non-stochastic and stochastic factors due to additional components not included in the fitting function (e.g., spiral arms, star-forming regions), asymmetries, shape mismatch, flat-fielding errors and so on. These non-random factors usually dominate the uncertainty of the parameters, and the uncertainties inferred from the covariance matrices are only lower-limit estimates \citep{Peng10}. Therefore, if we rely on the errors reported by GALFIT the uncertainties in the structural parameters of the BCGs could be severely underestimated. Indeed, these formal errors seem unrealistically small: typical GALFIT uncertainties for \re\ and $n$ are only $\sim1$--$2\%$. A more robust and realistic way of determining these uncertainties is clearly needed. 

We have measured the structural parameters of the BCGs in our sample using the SDSS $r$-band images. Independent measurements can also be obtained using the SDSS $g$-band images. In principle, the structural parameters could be wavelength-dependent. However, the $g-r$ colours of massive early-type galaxies with old stellar populations are quite spatially uniform and do not change much from galaxy-to-galaxy \citep[e.g.,][]{fukugita95}. Furthermore, morphological $k$-corrections are negligible for early-type galaxies between these two bands (e.g., \citealt{TaylorMager07}), so it is reasonable to expect that the intrinsic structural parameters will not change much between $g$ and $r$ band. Therefore, any differences in the measured parameters between these two bands should be largely dominated by measurement errors. Moreover, if there are significant wavelength-dependent differences in the measured parameters that are driven by real physical differences, it is reasonable to expect that these may correlate with other galaxy properties such as their colour, morphology, redshift, cluster velocity dispersion, etc. No such correlations were found, so we are confident that the intrinsic differences are not significant in these two bands.

We use GALAPAGOS to fit the SDSS $g$-band images of the BCGs in our sample in exactly the same way as we did for the $r$-band images. Figure~\ref{fig:error} shows a comparison of the \re\ and \rffs\ values obtained in both bands. Similar comparisons were carried out for the rest of the structural parameters. The scatter around the 1-to-1 relations is due, in principle, to both intrinsic wavelength-dependent differences and measurement errors. Since, as we have argued, the intrinsic differences are not expected to be significant between these two bands, the measurement errors should dominate the scatter. We can thus use this scatter as an estimate of realistic, albeit perhaps marginally pessimistic, parameter uncertainties. The average errors are $\delta (n) \simeq 0.9$, $\delta(\log r_{\rm e}) \simeq 0.16$, and $\delta(\log RFF_{1c}) \simeq 0.13$.

The right-hand panel of Figure~\ref{fig:error} shows that the errors in  \re\ and \reff\ are not correlated. This is an important point since these two are the main parameters that we will use as diagnostics in our analysis in Section~\ref{sec:border}.

\subsection{Single \sersic\ Models}
\label{sec:1cresult}
We analyse now the behaviour of four parameters derived from the best-fitting single-\sersic\ models along with the morphological classifications. Two of them, the \sersic\ index $n$ and the effective radius \re, provide information on the intrinsic properties of the BCGs. The other two, \reff\ and \chisq, show how well the models fit the real light distribution of the BCGs and also provide information about their detailed structure. The values of these parameters are listed in Appendix~A. Figure~\ref{fig:paradistr} shows the distribution of these parameters for the three main BCG morphologies. The  $\sigma$ value in each panel indicates the significance (confidence level) of the observed differences between the cD and elliptical BCG parameter distributions. These are derived from two-sample Kolmogorov-Smirnov tests.

\begin{figure}
\raggedright{\includegraphics[scale=0.51, angle=0]{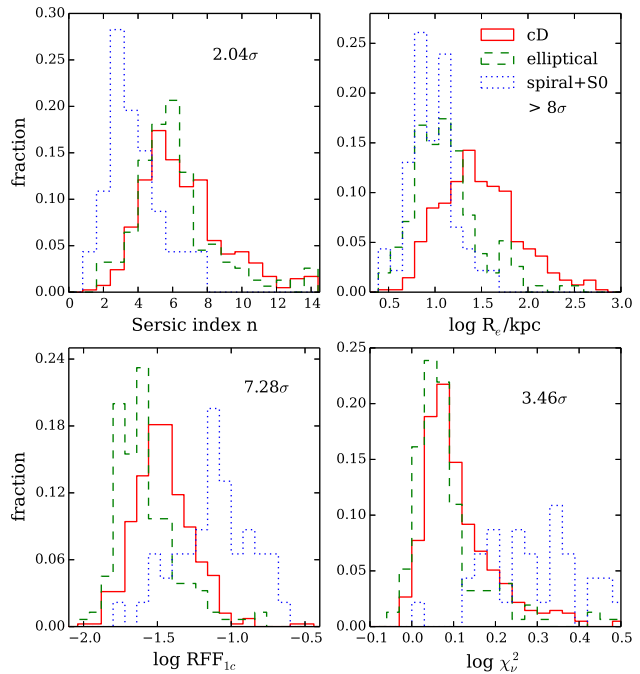}}
\caption{Distribution of the \sersic\ index $n$ (upper left), effective radius \re\ (upper right), $\log$\rffs\ (lower left) and $\log$\chisq\ (lower right) from single \sersic\ fits for the BCGs divided by morphology. The red solid line corresponds to cD galaxies, the green dashed line to ellipticals, and the blue dotted line to spirals and S0s. The  $\sigma$ value in each panel indicates the significance (confidence level) of the observed differences between the cD and elliptical BCG parameter distributions. These are derived from two-sample Kolmogorov-Smirnov tests.
\label{fig:paradistr}}
\end{figure}

\subsubsection{\sersic \ Index $n$}
\label{sec:1cn}

The \sersic\ index $n$ measures the concentration of the light profile, with larger $n$ corresponding to higher concentration. The upper left panel of Figure~\ref{fig:paradistr} presents the $n$ distributions for the three main BCG morphologies. It is clear that disk (spiral and S0) BCGs tend to have smaller values of $n$, as expected. However, the $n$ distribution for disk BCGs is skewed towards larger values ($n\gtrsim3$) than those of the normal disk galaxy population \citep[e.g., $n=2.5$ in][]{Shen03}. This is because most disk BCGs are early-type bulge-dominated spirals and S0s. Elliptical and cD BCGs tend to have larger $n$ values ($n\ge4$). The $n$ distributions of cD and elliptical BCGs are quite similar. A K--S test indicates that the distributions are not significantly different: the significance of any possible difference is just $2.04\sigma$.  

\subsubsection{Effective Radius \re }
The effective radius \re\ is a measurement of the extent (or size) of the light distribution. The upper right panel of Figure~\ref{fig:paradistr} shows the distributions of $\log R_{\rm e}$. Disk BCGs tend to have relatively small sizes, and the vast majority of them ($\sim 85\%$) have \re\ smaller than $\sim 15\,$\reunit. About $75\%$ of the elliptical BCGs also have $R_{\rm e} \lesssim 15\,$\reunit, while cD galaxies tend to be significantly larger. More than $60\%$ of cDs have $R_{\rm e} \gtrsim 15\,$\reunit. A K--S test demonstrates that the difference in \re\ distributions between cD and elliptical BCGs is very significant. This suggests that \re\ could be a good discriminator to separate cD and elliptical BCGs.

\subsubsection{Residual Flux Fraction and Reduced $\chi^2$}
\label{sec:rff1c}
The lower left panel of Figure~\ref{fig:paradistr} presents the \rffs\  distributions in a $\log_{10}$ scale, where \rffs\ denotes \reff\ for one-component models. The \rffs\ of disk BCGs has a much broader distribution and reaches significantly larger values than those of cDs and ellipticals. This reflects the fact that a single-\sersic\ model is not a good representation of the light distribution of galaxies with clear disks, spiral arms and star-forming regions. Early-type BCGs have smoother light distributions that can be reasonably well reproduced with a \sersic\ profile, and their \rffs\ tend to be smaller. However, there are statistically significant differences between the \rffs\ distributions of cD and elliptical BCGs. About $60\%$ of elliptical BCGs have \rffs\ values in the range corresponding to good fits (see Section~\ref{sec:1cvs2c} and Figure~\ref{fig:comp12BCG}), while just $\sim 25\%$ of cD galaxies do. This suggests that most elliptical BCGs can be well represented by single \sersic\ models, while most cD galaxies are harder to model with such a simple profile. Since an extended envelope is a general property of cD galaxies, their deviation from a single \sersic\ profile may be due, at least partially, to this extended envelope. This suggests that an additional model component may be required for them. We will re-visit two-component models in Section~\ref{sec:2cresult}. The clear difference in \reff\ suggests that \reff\ could be another good discriminator to separate cD and elliptical BCGs.

Similar conclusions can be reached from the the distributions of \chisq\ shown in the lower right panel of Figure~\ref{fig:paradistr}, albeit less clearly. This is not surprising since, as shown in Section~\ref{sec:1cvs2c}, both \reff\ and \chisq\ measure the strength of the residuals, but \chisq\ is significantly less sensitive. Therefore, \reff\ is expected to be more efficient for separating cD and elliptical BCGs than \chisq.\\

\noindent These results show a clear link between the visual morphologies of BCGs and their structural properties. Although cD galaxies tend to have similar shapes to elliptical BCGs, they usually have larger sizes and their structures generally deviate more from single \sersic\ profiles. In contrast, elliptical BCGs tend to be smaller, and their light profiles are statistically more consistent with single \sersic\ models. These structural differences, especially in \re\ and \reff, could therefore provide quantitative ways to separate elliptical and cD BCGs without relying on visual inspection. We will explore these issues in Section~\ref{sec:border}.

\subsection{\sersic+Exponential Models}
\label{sec:2cresult}

\begin{figure}
\raggedright{\includegraphics[scale=0.48]{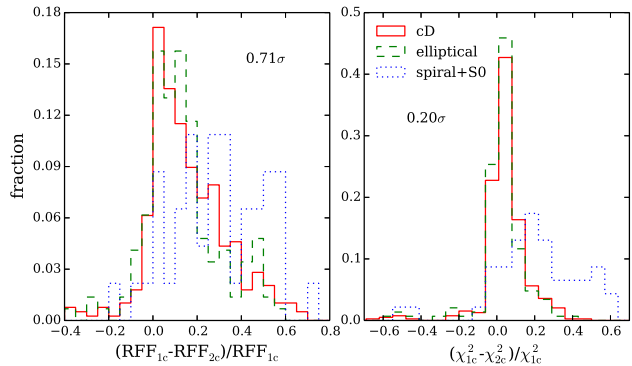}}
\caption{Comparison of the residuals between single \sersic\ and \sersic+Exponential models. The left panel shows the fractional differences in \reff\ obtained with two-component and one-component fits for cD (red solid line), elliptical (green dashed line), and disk (blue dotted line) BCGs. The right panel shows the corresponding fractional differences for \chisq . 
\label{fig:RFF12c}}
\end{figure}

The \reff\ distributions shown in Section~\ref{sec:1cresult} indicate that elliptical BCGs are statistically better fitted by a single \sersic\ model than cDs. Since a distinctive feature of cD galaxies is their extended luminous halo, two-component models may be more appropriate to describe accurately the light distributions of cD BCGs. Following \cite{Seigar07} and \cite{Donzelli11}, we explore here how a model consisting of an inner \sersic\ profile and an outer exponential envelope performs when fitting BCG images. The fitting process was described in detail in Section~\ref{sec:BDfit}.

As shown in Section~\ref{sec:1cvs2c}, both \reff\ and \chisq\ can provide quantitative information to assess whether BCGs are better fitted by a two-component model than by a one-component model, at least in very clear cases. Figure~\ref{fig:RFF12c} shows a comparison of these parameters obtained for single \sersic\ and \sersic+Exponential models. In the left panel we show a histogram of the fractional differences in the \reff\ values $(\rffsx-\rffbdx)/\rffsx$ for all three BCG types. The right panel shows the corresponding \chisq\ fractional differences $(\chisx-\chibdx)/\chisx$. It is clear that for disk BCGs, the \sersic+Exponential model does a better job. This is not surprising since spiral and lenticular galaxies contain clearly distinct bulges and disks. For elliptical BCGs the improvement in \reff\ and \chisq\ for two-component models is generally quite small, as expected: elliptical galaxies are known to be reasonably well fitted by \sersic\ models, so the extra component does not improve the residuals significantly. Perhaps surprisingly, the improvement is also only marginally better for cDs: the typical fractional differences for cD galaxies are $(\rffsx-\rffbdx)/\rffsx=0.11^{+0.14}_{-0.08}$ and $(\chisx-\chibdx)/\chisx=0.035^{+0.053}_{-0.029}$ (median $+$/$-$ 1st and 3rd quartiles of the parameter distributions).

Since the distributions shown in Figure~\ref{fig:RFF12c} for ellipticals and cDs are statistically indistinguishable, there is no clear separation that could be used to distinguish elliptical and cD BCGs by comparing one-component and two-component fits. Moreover, on average, \sersic+Exponential model does not fit the profile of cD BCGs clearly better than single \sersic\ model. The reason is that for cD BCGs the values of \reff\ and \chisq\ are generally not dominated by the presence or absence of a second exponential model component but by other structures present in the residual images, such as double cores. Since there is no clear improvement in the \sersic+Exponential model, the model with the smallest number of parameters (i.e., single \sersic\ model) will be preferred for simplicity. The following discussions are based on the results from the single \sersic\ fits.

\subsection{Summary of Section~\ref{sec:structureresult}}
\label{sec:1cdiscuss}

In this section we have analysed the differences in the structural properties of BCGs as a function of morphology. These structural parameters have been derived from one-component (\sersic) and two-component (\sersic+exponential) model fits. Disk BCGs (a small minority) have smaller \sersic\ indices ($n$) than elliptical and cD BCGs, as expected. They also have different, generally broader, distributions of \reff\ and \chisq. Elliptical and cD BCGs have similar $n$ values, but cDs tend to have larger values of \re, \reff\ and \chisq. These differences do not depend strongly on whether we use one- or two-component models. 

The observed structural differences could provide quantitative ways to separate elliptical and cD BCGs without relying on visual inspection. We explore these in section~\ref{sec:border}. Furthermore, the differences we have found in the structural parameters suggest that the formation histories of elliptical and cD BCGs may be different. For instance, gas-rich major mergers and other dissipative processes may be responsible for building the inner (\sersic-like) component, while dissipationless minor mergers may contribute to the build-up of the outer extended envelope and to the growth of galaxy sizes (e.g., \citealt{Oser10}; \citealt{Johansson12}; \citealt{Huang13}). We will explore in a subsequent paper (Zhao et al., in preparation) whether the morphological and structural properties of BCGs are linked to other intrinsic BCG properties such as their stellar mass, and/or to the properties of their environment. These links will provide more clues to the formation history of cDs/BCGs.

\section{Separating elliptical and cD BCGs}
\label{sec:border}

The results of Section~\ref{sec:1cresult} suggest that we may be able to use the different distributions of cD and non-cD BCGs on the $\log$\re--$\log$\rffs\ plane to separate them in an objective, quantitative and automatic way. Figure~\ref{fig:Abeta125} shows that cDs are clearly segregated from other BCGs in this two-dimensional parameter space. We attempt to find a robust, well-defined way to separate, statistically, cD and non-cD BCGs using the information provided by this diagram. In other words, we suppose to find an ``optimal border'' that can separate them.

\subsection{Method Description and the Optimal Border}
\label{sec:bestborder}
Ideally, any process that selects cD galaxies from a sample of BCGs needs to have high completeness (i.e., select as many of the cDs present in the sample as possible), while avoiding contamination from non-cDs (i.e., maximising the purity of the sample). These two requirements compete with each other, and increasing completeness often results in a decrease in sample purity, and vice-versa. We need therefore to find the best compromise between these competing requirements. In general, the optimal solution will depend on the specific intent for the selected sample, and therefore on the decision of how much weight to give to completeness and to purity. It is useful to define a measurement on the quality of the selection method that combines both requirements in a well-defined way. The optimal solution will then be obtained by maximising this quality parameter.

Following \cite{Hoyos11} the \textit{sensitivity}, which is often known as \textit{completeness} in astronomy, is defined as:
\begin{equation}
r=\dfrac{\rm \# True Positives}{\rm \# True Positives+ \# False Negatives}.
\end{equation}
Similarly, we define \textit{specificity} as:
\begin{equation}
p=\dfrac{\rm \# True Negatives}{\rm \# True Negatives+ \# False Positives}.
\end{equation}
A ``True Positive'' is an object retrieved by the selection process with the required properties (i.e., a cD galaxy that is correctly selected as such). A ``False Negative'' is an item that is not retrieved by the selection process but does present the needed properties (a cD galaxy that is not selected). A ``True Negative'' is an item that is rightfully rejected by the selection process since it does not have the required properties (for instance, an elliptical galaxy that is not selected as a cD). A ``False Positive'' is an item that is incorrectly picked up by the selection process, but does not have the properties of interest (for example, an elliptical galaxy that is wrongly selected as a cD). 

\textit{Sensitivity} and \textit{specificity} can be combined into a single number, known as the \fscore\ (\citealt{vanRijsbergen79}), which provides a single measure on the quality of the selection process.  The \fscore\ is just a weighted harmonic average of $r$ and $p$, 
\begin{equation}
F_\beta=\dfrac{(1+\beta ^2)\times p \times t}{\beta ^2 \times p+r},
\end{equation}
where $\beta$ is a control parameter that regulates the relative importance of completeness with respect to specificity. This is a user-supplied value that depends on the particular goals of the study. We will explore later how the choice of $\beta$ affects our selecting results. At this stage, a value of $\beta= 1.25$ is used, which can be thought of as weighing completeness more than the lack of contamination. For our BCG samples, the \fscore\ is used to grade the performance of the diagnostics we use when separating cD galaxies from the parent population. 

\begin{figure}
\raggedright{\includegraphics[scale=0.4, angle=0]{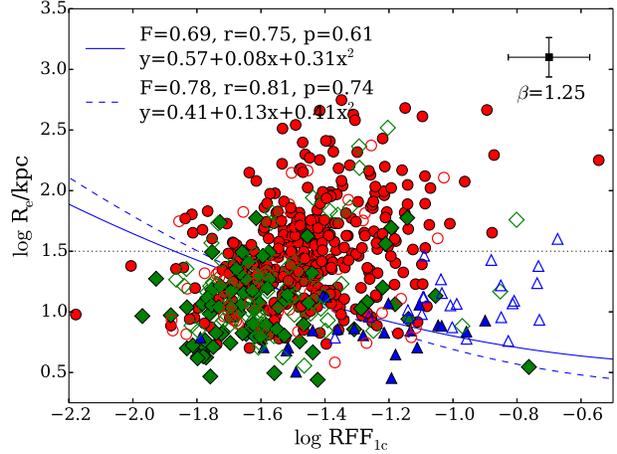}}
\caption{$\log \rex$ vs.\ $\log\rffsx$ for the BCGs in our sample. We use this diagram to find the optimal border to separate cD from non-cD BCGs. The symbols are the same as in Figure~\ref{fig:skyparam}. The black dotted line is the ``first guess'' for the border. The blue solid curve is the optimal border determined when we consider all cD BCGs (cD, cD/E and cD/S0) as cD galaxies. The blue dashed curve is the optimal border determined when we consider only pure cD and pure elliptical BCGs (excluding all cD/E, cD/S0, E/cD, E/S0, spiral and S0 BCGs). The legend shows the maximum \fscore\ for the optimal borders and the corresponding completeness $r$ and specificity $p$. The equations defining the optimal borders are also shown. The error bar shows the mean error of each parameter. We used $\beta=1.25$ in this case.  
\label{fig:Abeta125}}
\end{figure}

The selection process that we will apply to the parent population of BCGs in order to select cD galaxies will be defined by a ``border'' in the $\log$\re--$\log$\rffs\ plane (see Figure~\ref{fig:Abeta125}). This border will be represented by a second-order polynomial in the horizontal coordinate. Higher-order polynomials (or more conplex functions) could be used, but the additional complexity is not required here. In our specific problem, the cD galaxies play the role of the ``items presenting the required properties'' discussed above, and the parent population is the complete sample of BCGs.

According to the definition of \textit{sensitivity} and \textit{specificity}, the BCGs in the parent sample are  classified into four categories by their position relative to the border. In the $\log$\re--$\log$\rffs\ plane, cD galaxies dominate the region of large \re\ and \rffs. We therefore define this  region as the ``cD side''. Thus
\begin{itemize}
\item[$\bullet$] cD galaxies that fall on the cD side of the border are True Positives.
\item[$\bullet$] cD galaxies that do not fall on the cD side of the border are called False Negatives.
\item[$\bullet$] elliptical and disk (spiral and S0) BCGs that fall on the cD side are regarded as False Positives.
\item[$\bullet$] elliptical and disk (spiral and S0) BCGs that do not fall on the cD side of the border are True Negatives.
\end{itemize}
The optimal border is found by maximising the \fscore\ value. Following the method described in \cite{Hoyos11}, we use the Amoeba algorithm \citep{Press88} to carry out this maximization and find the polynomial defining the border.

It is clear from  Figure~\ref{fig:Abeta125} that the selected galaxy sample on the cD side of the optimal border will not contain only cD galaxies, and a degree of contamination will be present. We define contamination \citep{Hoyos11} as:
\begin{equation}
C=\dfrac{\textrm{\#non-cDs tested as positive}}{\textrm{\#all positives}}.  
\end{equation}
The numerator are the non-cD BCGs which are on the cD side of the optimal border. The denominator of this fraction includes both cD galaxies and non-cD BCGs on the cD side.

\begin{figure}
\raggedright{\includegraphics[scale=0.4, angle=0]{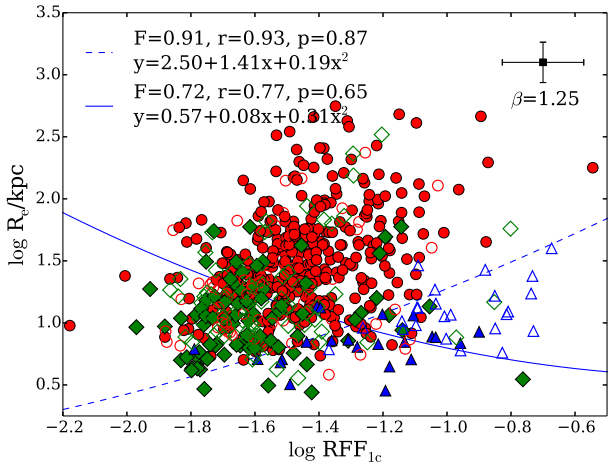}}
\caption{Two-step process to select cD BCGs. Symbols and legend are the same as in Figure~\ref{fig:Abeta125}. Disk (spiral and S0) BCGs are separated from non-disk BCGs (cDs and ellipticals) first using the \bestb\ shown as the blue dashed curve. cD galaxies are then selected using the \bestb\ shown as the blue solid curve. See text for details.   
\label{fig:twoborder}}
\end{figure}

Figure~\ref{fig:Abeta125} shows the $\log\rex$--$\log\rffsx$ plane for the BCGs in our sample. The Amoeba algorithm requires a first guess for the border, shown by the black horizontal dotted line. The optimal border determined by the algorithm does not depend on the exact initial guess. The blue solid curve is the optimal border determined when we consider all cD galaxies (cD, cD/E and cD/S0) as cD galaxies. This border, computed using $\beta=1.25$, has \fscore$\,=0.69$.  $75\%$ of all the cD galaxies are above the border ($r=0.75$), and thus selected from the parent sample. The remaining $25\%$ are mixed with the elliptical and disk BCGs in the region below the border. This selection therefore yields 75\% completeness. The galaxy sample above the border contains $311$ cD galaxies and $79$ non-cD BCGs resulting in a $\sim 20\%$ contamination in the selected cD samples. In the region below the border there are $103$ cD galaxies and $122$ and non-cD BCGs. Thus, the non-cD BGC sample has a contamination of $46\%$ from cD galaxies. This indicates that this technique is more effective (cleaner) at selecting cD galaxies than at selecting non-cD BCGs. 

Note that if we consider a ``cleaner'' sample that contains only pure cD and pure elliptical BCGs (excluding all cD/E, cD/S0, E/cD, E/S0, spiral and S0 BCGs), the optimal border (blue dashed curve in Figure~\ref{fig:Abeta125}) does not change significantly, but the quality of the selection as determined by the \fscore\ value, the completeness $r$ and the specificity $p$ improves. This is not surprising: the identification of BCGs as pure cDs/Es (as opposed to the ``dubious'' ones) depends on more secure morphological characteristics which should be linked more clearly to the structural parameters. However, considering only this cleaner sample is not a realistic scenario since in practical cases we would like to start from a full sample of BCGs and find which ones are cDs. Nevertheless, it is reassuring that the border we determine does not depend very strongly on the exact training set used.

On the selected cD side, spiral BCGs are an important source of contamination. However, since most of them appear in the large \rffs\ region, it would be possible to go a step further to implement a simple further refinement in our method to separate spirals from the selected cDs: very few cD galaxies have $\log \rffsx$ larger than $\sim -1.1$. This would significantly improve the purity of the cD sample at very little cost in terms of its completeness. 

Moreover, it is clear from Figure~\ref{fig:Abeta125} that all disk BCGs (spirals and S0s) contribute significantly to the contamination of either the cD or the elliptical samples separated by the best border. However, we can use the fact that disk BCGs distribute over a distinct area on the $\log\rex$--$\log\rffsx$ plane to apply a two-step process to exclude them from our cD selection. First, the disk BCGs can be separated from the elliptical and cD BCGs, and then the cD BCGs can be selected out of the rest BCG sample. Figure~\ref{fig:twoborder} illustrates the results of this two-step selection. The blue dashed curve is the \bestb\ determined in the the first step. By excluding disk BCGs using this border, a very complete ($r=0.93$) and pure ($p=0.87$) non-disk BCG sample is built. The cDs can then be separated from the ellipticals using the \bestb\ shown by the blue solid curve with a completeness of $77\%$ ($305$ cDs are selected), and a contamination of only $14\%$. Compared to the single-step cD selection ($311$ cDs were selected with $20\%$ contamination), the two-step process clearly selects a very similar number of cDs but with better purity. The decision on whether the increase in purity is worth the additional complexity is left to the reader. In the reminder of this paper we will use the single-step selection process for simplicity.

The automatic techniques we have developed can be applied to any BCG sample, but the \bestb\ needs to be adapted and calibrated using the imaging data from which the parent sample was derived. The calibration can be performed using a sub-sample of visually-classified BCGs, and then automatically applied to the complete sample using the structural parameters determined from standard single-\sersic\ fits. 

A $\beta$ value needs to be chosen depending on whether we are more interested in the completeness of the cD sample or in its purity, but we suggest that $\beta=1.25$ represents a reasonable compromise (see section~\ref{sec:betaparameter}). Furthermore, it is important to remember that this method works better at selecting a sample of cD galaxies rather than a sample of non-cDs.

\subsection{Distance to the Optimal Border}
\label{sec:distancetoborder}

\begin{figure}
\centering
\centering{\includegraphics[scale=0.55]{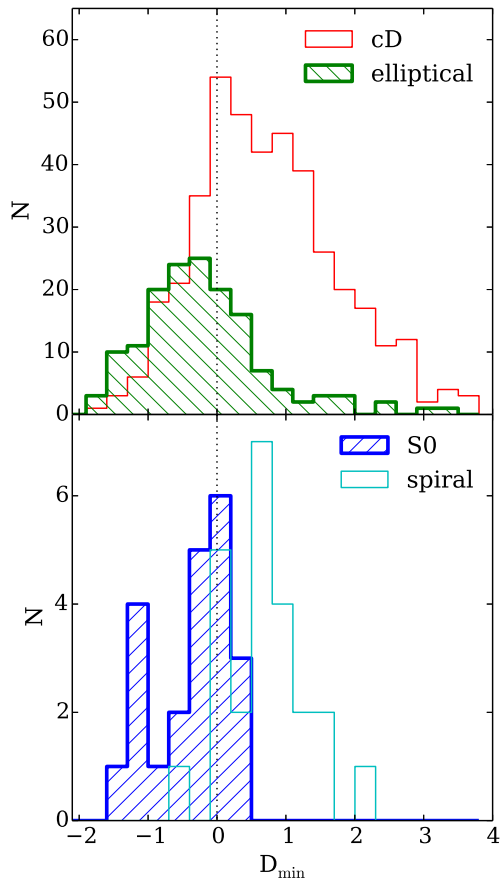}}  
\caption{Distribution of the minimum distances to the \bestb\ shown in Figure~\ref{fig:Abeta125} for the cD and elliptical BCGs (top panel) and the spiral and S0 BCGs (bottom panel). Positive and negative distances correspond to points above and below the \bestb\ line respectively.}
\label{fig:borderdistanceall}
\end{figure}

\begin{figure}
\centering
\centering{\includegraphics[scale=0.55]{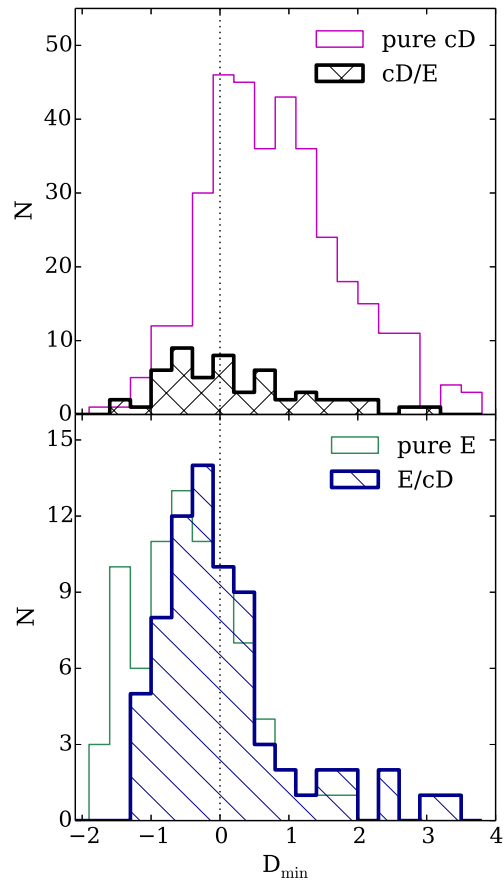}}  
\caption{Distribution of the minimum distances to the \bestb\ shown in Figure~\ref{fig:Abeta125} for the pure cD BCGs and cD/E BCGs (top panel). The bottom panel shows the corresponding histograms for pure E BCGs and E/cD BCGs.}
\label{fig:borderdistamcecDs}
\end{figure}

It is informative to explore the distribution of the points in the $\log \rex$--$\log \rffsx$ plane (Figure~\ref{fig:Abeta125}) in terms of their minimum (perpendicular) distance to the \bestb. We define the distance from each point to the \bestb\ as
\begin{equation}
D=\sqrt{\left( \frac{\Delta \log \rffsx}{\sigma_{\log \rffsx}}\right) ^2  + \left( \frac{\Delta \log \rex}{\sigma_{\log \rex}}\right) ^2},
\end{equation}
where $\Delta \log \rffsx$ is the difference in $\log \rffsx$ between the data point and the \bestb, and $\sigma_{\log \rffsx}$ is the dispersion in $\log \rffsx$ computed for all the points. $\Delta \log \rex$ and $\sigma_{\log \rex}$ have a similar meaning but for $\log \rex$. Note that, because the units of the $x$ and $y$ axes are different, the distance is measured in units of the scatter of each parameter.  For each point, the minimum distance $D_{\rm min}$ can be then determined. Figure~\ref{fig:borderdistanceall} shows the distribution of these minimum distances for the different morphologies. As expected, the vast majority ($>80\%$) of the cDs show positive distances (they are above the \bestb\ line) while most of the ellipticals have negative ones. Under $20\%$ of the cDs spill over to the negative region, severely contaminating the non-cD sample, while a few ellipticals weakly contaminate the cD region. The measurement errors in $\log \rex$ ($\sim0.16$) and $\log \rffsx$ ($\sim0.13$) result in distance errors on the order of $0.7$ in this metric. This contributes to the cDs' ``spillover'', but does not completely explain it. Reducing the measurement errors would certainly improve the performance of our method, but it would never make it perfect.

Interestingly, the spiral and S0 BCGs are quite well separated: the former show mostly positive distances while the later have mostly negative ones. This is mainly due to spirals having generally larger \rffs\ values because the spiral arms and star-forming regions are not included in the \sersic\ models, while the S0s are smoother. This clear separation provides a possible way to separate spiral and S0 galaxies, but this needs to be further tested with large disk samples.

Another interesting result is that BCGs classified as pure and uncertain cDs (e.g., cD/E) have very different minimum distance distributions (Figure~\ref{fig:borderdistamcecDs}, top panel). About half of the cD/E BCGs have negative distances (i.e., are on the wrong side of the border), but only $\simeq20$\% of the pure cDs do. Most of the spillover of the pure cDs into the negative region, however, can be explained by the measurement errors. It should be noticed that the difficulties inherit in the visual morphological classification are directly reflected in the structural parameters: when the visual classifier is certain that a BCG is a cD, its structural parameters almost always confirm it, while in uncertain cases (e.g., cD/E) the structural parameters reflect this uncertainty. Similar conclusions can also be obtained from the pure elliptical BCGs and uncertain ones (e.g., E/cD), as shown in the bottom panel of Figure~\ref{fig:borderdistamcecDs}.

This analysis confirms the visual impression in terms of the BCG structure that there is a continuous distribution in the properties of the BCG extended envelopes, ranging from undetected (pure E class) to clearly detected (pure cD class), with the intermediate classes (E/cD and cD/E) showing increasing degrees of envelope presence. This continuous distribution in envelope detectability is reflected quantitatively in the structural parameters of the BCGs, by the minimum distance to the \bestb\ providing some indication of the relative importance of the envelope.

\subsection{Effect of the $\beta$ Parameter}
\label{sec:betaparameter}
\begin{figure}
\subfigure{
\raggedright{\includegraphics[scale=0.4]{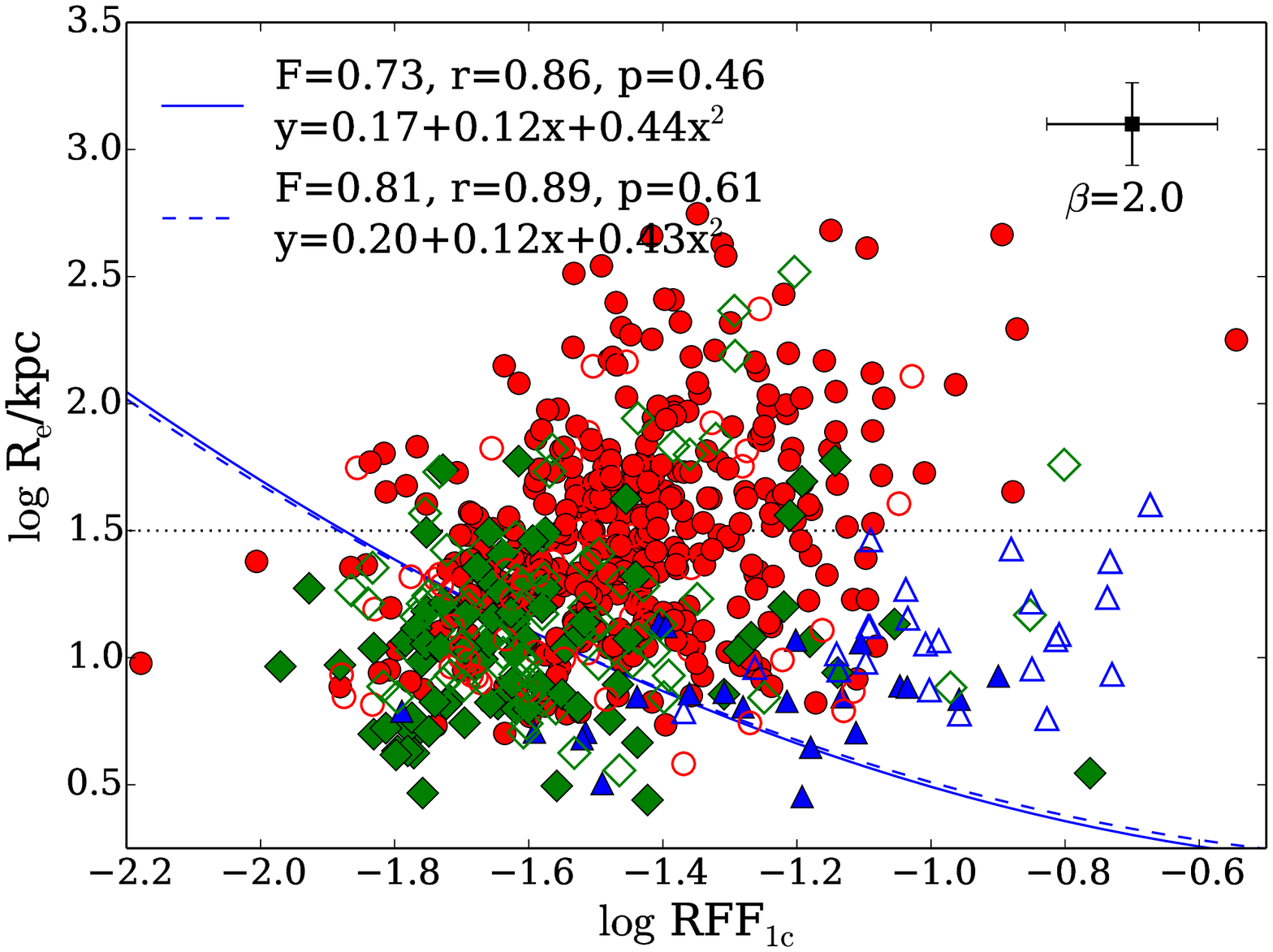}}}  
\subfigure{
\raggedright{\includegraphics[scale=0.4]{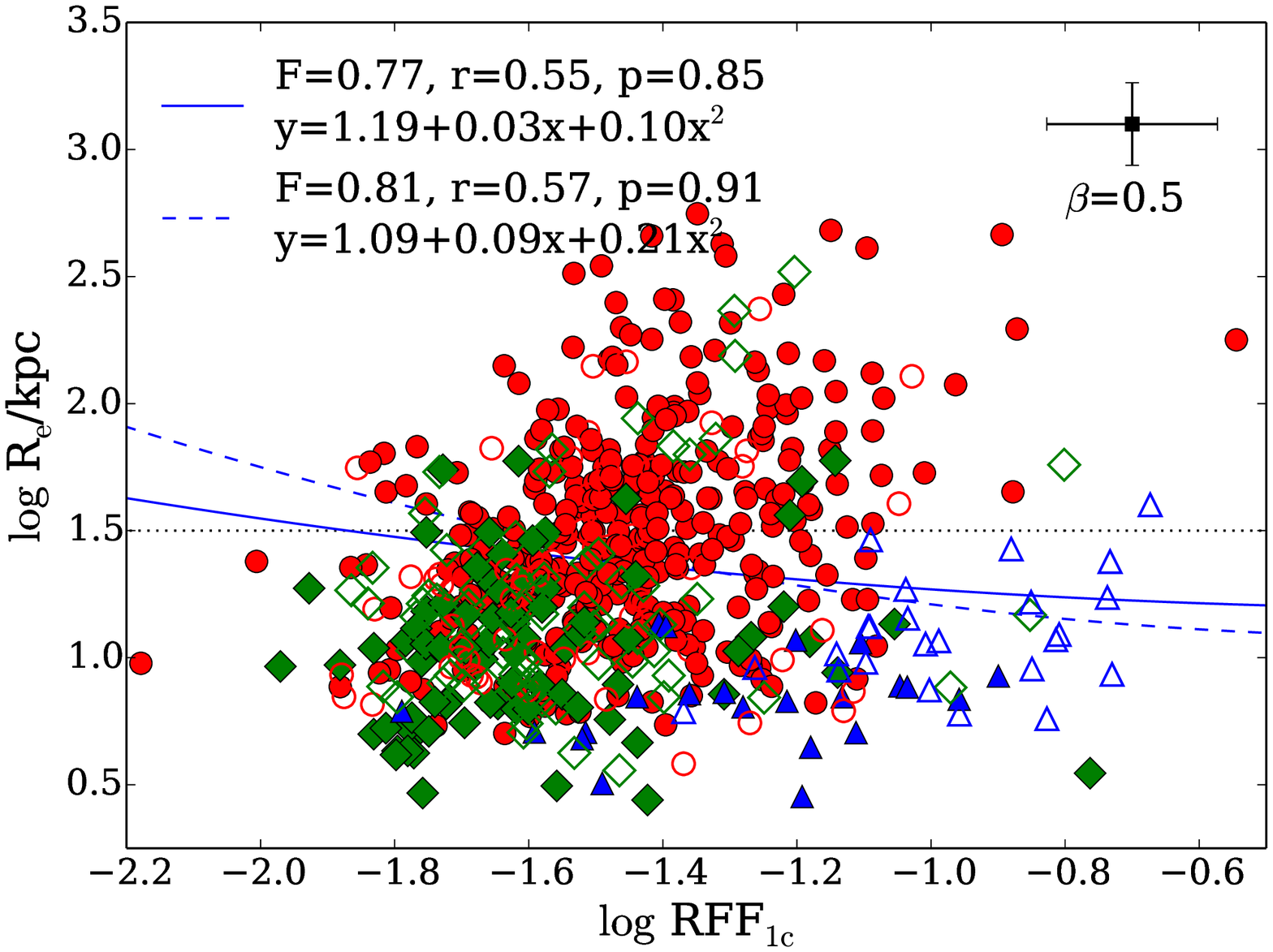}}}
\caption{Illustration of the effect of $\beta$ on the optimal border. The symbols, lines and legends have the same meaning as in Figure~\ref{fig:Abeta125} but we use $\beta=2.0$ for the upper panel and $\beta=0.5$ for the lower panel. With $\beta=2.0$ we give more weight to the completeness than to  the lack of contamination. When using $\beta=0.5$, the lack of contamination is given more importance than achieving higher completeness.  The choice on $\beta$ depends on the aims of the specific research.}
\label{fig:Abeta0520}
\end{figure}

In the \fscore\ definition, the $\beta$ parameter is used to apportion weight to the completeness and the specificity. For larger values of $\beta$ the completeness is given a larger weight than the lack of contamination.  Conversely, smaller values of $\beta$ prioritise lack of contamination above completeness. To test how changing $\beta$ affects the results of the selection process, we repeat the exercise carried out in Section~\ref{sec:bestborder} but using $\beta=2.0$ and $\beta=0.5$ in the determination of the optimal border.

Figure~\ref{fig:Abeta0520} shows the \bestb\ for $\beta=2.0$ (upper panel) and $\beta=0.5$ (lower panel). It is clear that the $\beta$ parameter has a decisive impact on the selection of potential cD galaxies. As shown in the upper panel, when compared to the $\beta=1.25$ results, $11\%$ more galaxies are correctly identified as cDs, significantly increasing the completeness. The price paid is that the specificity goes down from $61\%$ to $46\%$ since more non-cD BCGs are included. Conversely, in the lower panel ($\beta=0.5$) the selected cD sample is purer ($p=0.85$), but at the expense of completeness, with $20\%$ fewer cD galaxies selected when compared with the $\beta=1.25$ result.

With $\beta=2.0$, the contamination of the cD sample by non-cDs is $23\%$, while the contamination of the non-cD sample by cDs is $39\%$. With $\beta=0.5$, the corresponding values are $12\%$ and $52\%$ respectively. Therefore, for any value of $\beta$ this selecting technique is cleaner and more effective at selecting cD galaxies than at selecting non-cD BCGs.

As before, if we consider a cleaner sample that contains only pure cD and pure elliptical BCGs, the optimal border (blue dashed curve) does not change significantly, but the \fscore\ value, the completeness $r$ and the specificity $p$ improve. However, we have argued that this does not represent a realistic scenario.

We conclude that $\beta=1.25$ represents a good compromise, as its \bestb\ picks up a cD galaxy sample reasonably complete, and with relatively small contamination. However, no single value of $\beta$ can be considered to be ``correct'' and needs to be set according to the scientific goals of the study.

\section{Conclusions}
\label{sec:conclude}

In this paper we have analysed a well-defined sample of $625$ low-redshift Brightest Cluster Galaxies published in \citet{Linden07} with the aim of linking their morphologies to their structural properties. We morphologically classified the BCGs using SDSS $r$-band images and found that over half of them ($\sim57$\%) are pure cD galaxies and pure elliptical BCGs constitute $\sim13$\% of the sample. The intermediate classes (mostly cD/E or E/cD) account for $\sim21$\%. It suggests a continuous distribution in the properties of the BCG extended envelopes, ranging from undetected (pure E class) to clearly detected (pure cD class), with the intermediate classes (E/cD and cD/E) showing increasing degrees of envelope presence. We found this continuous distribution in envelope detectability is reflected quantitatively in the structural parameters of the BCGs. There is also a minority of BCGs that are neither cD nor elliptical. About $7$\% are disk galaxies (spirals and S0s, in similar proportions) and the rest ($\sim2$\%) are in merging (see appendix~A).

In order to link the morphologies of the BCGs to their structural properties, we have fitted the BCGs light distributions with the SDSS $r$-band images using one-component (\sersic) and two-component (\sersic+Exponential) models. We first characterised how well the models fit the target BCG by using two quantitative diagnostics. One diagnostic is the residual flux fraction (\reff), which measures the fraction of the galaxy flux presenting in the residual images after subtracting the models. The other diagnostic is the reduced \chisq. We concluded that generally it is very difficult to find a robust diagnostic to decide, in a statistic way, whether a one-component or a two-component model is preferred for BCGs, especially for cD galaxies. Since there is no evident improvement by using two-component model fits, our other conclusions rely on the one-component \sersic\ fits.

From simple one-component \sersic\ profile fits, we have found a clear link between the BCGs morphologies and their structures, and claimed that a combination of the best-fit parameters can be used to separate cD galaxies from non-cD BCGs. In particular, cDs and non-cDs show very different distributions in the \re--\rffs\ plane, where \re\ is the effective radius and \rffs\ is the residual flux fraction, both determined from \sersic\ fits. cDs have, generally, larger \re\ and \rffs\ values than ellipticals. Therefore we found, in a statistically robust way, a boundary to separate cD and non-cD BCGs in this parameter space. BCGs with cD morphology can be selected with reasonably high completeness ($\sim 75\%$) and low contamination ($\sim 20\%$).

This automatic and objective technique can be applied to any current or future BCG samples which have good quality images. The method needs to be adapted and calibrated using the imaging data from which the parent sample was derived. Once calibrated with a representative sub-sample of visually-classified BCGs, this technique can be applied to the complete sample using the structural parameters determined from standard single-\sersic\ fits.

In a subsequent paper (Zhao et al., in preparation) we will explore how the morphological and structural properties of BCGs are linked to other intrinsic BCG properties such as their stellar mass, and/or to the properties of their environments. These links will provide more clues to the formation history of cDs/BCGs.

\section*{Acknowledgments}

DZ's work is supported by a Research Excellence Scholarship from the University of Nottingham and the China Scholarship Council. AAS and CJC acknowledge financial support from the UK Science and Technology Facilities Council. This paper is partially based on SDSS data. Funding for SDSS-III has been provided by the Alfred P. Sloan Foundation, the Participating Institutions, the National Science Foundation, and the U.S. Department of Energy Office of Science. The SDSS-III web site is http://www.sdss3.org/. SDSS-III is managed by the Astrophysical Research Consortium for the Participating Institutions of the SDSS-III Collaboration including the University of Arizona, the Brazilian Participation Group, Brookhaven National Laboratory, Carnegie Mellon University, University of Florida, the French Participation Group, the German Participation Group, Harvard University, the Instituto de Astrofisica de Canarias, the Michigan State/Notre Dame/JINA Participation Group, Johns Hopkins University, Lawrence Berkeley National Laboratory, Max Planck Institute for Astrophysics, Max Planck Institute for Extraterrestrial Physics, New Mexico State University, New York University, Ohio State University, Pennsylvania State University, University of Portsmouth, Princeton University, the Spanish Participation Group, University of Tokyo, University of Utah, Vanderbilt University, University of Virginia, University of Washington, and Yale University. 

\bibliographystyle{mnras}     
\bibliography{BCG_paper1}

\appendix

\section{Data Table}

 Table~A1 contains the main properties of the BCGs discussed in this paper. The full table is published electronically. 

\begin{landscape}
\begin{table} 
\centering
\label{tab:L07data}
\begin{tabular}{rrrrrrrrrrll}
\hline\hline
 ID2  &ID3  &RA &DEC  &$z$  &$\sigma_{\rm cl}$  & $\log R_{\rm e,1c}$  &$n_{\rm 1c}$  &$RFF_{\rm 1c}$  &$\chi^2_{\rm 1c}$  & Type  &Comments \\
    (1)  &  (2)   &deg (3)    &deg (4)    &  (5)  &km$\,$s$^{-1}$ (6)    &kpc (7)  & (8)  & (9)  & (10)  & (11) & (12) \\ [0.5ex]
\hline   
1011    & 1013    & 227.107346    & $-$0.266291    & 0.091    & 748    & 1.527    & 5.38    & 0.08190    & 1.752    & cD    & Clear halo; perhaps interacting  \\[0.5ex]   
1023    & 1025    & 153.409478    & $-$0.925413    & 0.045    & 790    & 1.908    & 6.25    & 0.05052    & 1.374    & cD    & Clear halo; interacting with fainter galaxies  \\[0.5ex]   
1064    & 1075    & 153.437067    & $-$0.120224    & 0.094    & 875    & 1.312    & 4.49    & 0.02648    & 1.086    & E/cD    &     \\[0.5ex]   
--    & 1027    & 191.926938    & $-$0.137254    & 0.088    & 1020    & 1.063    & 4.42    & 0.06594    & 1.903    & E    & Interacting/merging with bright early-type  \\[0.5ex]   
--    & 1389    & 202.337884    & 0.749685    & 0.080    & 853    & 1.044    & 6.02    & 0.01990    & 1.087    & E/cD    & Faint/small halo  \\[0.5ex]   
2040    & 2050    & 17.513187    & 13.978117    & 0.059    & 759    & 2.408    & 9.77    & 0.04122    & 1.224    & cD    & Several bright-ish companions  \\[0.5ex]   
1052    & 1058    & 195.719058    & $-$2.516350    & 0.083    & 749    & 1.627    & 4.89    & 0.04694    & 1.455    & cD    & Multiple merger  \\[0.5ex]   
1034    & 1036    & 192.308670    & $-$1.687394    & 0.085    & 771    & 0.977    & 4.86    & 0.02102    & 1.115    & E    &     \\[0.5ex]   
1041    & 1044    & 194.672887    & $-$1.761463    & 0.084    & 771    & 2.318    & 5.64    & 0.05023    & 1.280    & cD    & Very large, elongated halo; some faint companions  \\[0.5ex]   
--    & 1126    & 192.516071    & $-$1.540383    & 0.084    & 878    & 2.039    & 9.12    & 0.04520    & 1.348    & cD    & Interacting with faint companions  \\[0.5ex]   
3002    & 3004    & 258.120056    & 64.060761    & 0.080    & 1156    & 1.667    & 4.81    & 0.02561    & 0.991    & cD    &     \\[0.5ex]   
3096    & 3283    & 135.322540    & 58.279747    & 0.098    & 756    & 1.866    & 6.96    & 0.05535    & 1.144    & cD    & Merging with bright companion  \\[0.5ex]   
1045    & 1048    & 205.540176    & 2.227213    & 0.077    & 828    & 0.883    & 2.52    & 0.10689    & 11.280    & E/cD     & Multiple merger  \\[0.5ex]   
1003    & 1004    & 184.421356    & 3.655806    & 0.077    & 966    & 1.753    & 4.75    & 0.05233    & 1.225    & cD/E    & Interacting/merging with early-type  \\[0.5ex]   
--    & 1456    & 173.336242    & 2.199054    & 0.099    & 746    & 1.696    & 8.09    & 0.02573    & 1.128    & cD    &     \\[0.5ex]   
1053    & 1061    & 228.220703    & 4.514004    & 0.038    & 789    & 0.875    & 7.54    & 0.01749    & 1.074    & cD    &     \\[0.5ex]   
2163    & 2074    & 314.975446    & $-$7.260758    & 0.079    & 765    & 1.231    & 8.03    & 0.04481    & 1.315    & E/cD    &     \\[0.5ex]   
2002    & 2002    & 358.557007    & $-$10.419200    & 0.076    & 812    & 2.660    & 11.12    & 0.03832    & 1.201    & cD    & Many faint and bright-ish companions  \\[0.5ex]   
2006    & 2013    & 10.460272    & $-$9.303146    & 0.056    & 903    & 1.433    & 1.62    & 0.04140    & 1.477    & cD    & Several faint companions  \\[0.5ex]   
1355    & 1460    & 175.554108    & 5.251709    & 0.097    & 1074    & 0.952    & 5.30    & 0.01557    & 1.052    & cD    & Interacting with faint galaxy; faint but clear halo  \\[0.5ex]   
1058    & 1069    & 184.718166    & 5.245665    & 0.076    & 721    & 1.988    & 7.98    & 0.04144    & 1.251    & cD    & Interacting with faint galaxies  \\[0.5ex]   
1002    & 1002    & 159.777581    & 5.209775    & 0.069    & 800    & 1.740    & 8.40    & 0.03838    & 1.321    & cD/E    & Clear halo  \\[0.5ex]   
--    & 1276    & 183.271286    & 5.689677    & 0.081    & 729    & 0.995    & 5.30    & 0.02142    & 1.151    & E    &     \\[0.5ex]   
1039    & 1042    & 228.808792    & 4.386210    & 0.098    & 857    & 1.800    & 8.77    & 0.04365    & 1.205    & E/cD     & Some halo? faint companions  \\[0.5ex]   
--    & 3332    & 124.471428    & 40.726395    & 0.063    & 802    & 1.463    & 6.40    & 0.08125    & 2.639    & SB0    &     \\[0.5ex]   
3011    & 3028    & 204.034694    & 59.206401    & 0.070    & 872    & 2.120    & 7.86    & 0.08172    & 1.556    & cD    & Several faint companions  \\[0.5ex]   
1001    & 1001    & 208.276672    & 5.149740    & 0.079    & 746    & 1.820    & 7.85    & 0.02720    & 1.128    & E/cD    &     \\[0.5ex]   
3004    & 3012    & 255.677078    & 34.060024    & 0.099    & 1127    & 1.717    & 3.54    & 0.08433    & 1.949    & cD     & Late merger?  \\[0.5ex]   
--    & 3094    & 254.933115    & 32.615319    & 0.098    & 875    & 1.291    & 3.50    & 0.02878    & 1.069    & cD    & Very faint companions  \\[0.5ex]   
--    & 1066    & 202.795126    & $-$1.730259    & 0.085    & 814    & 1.942    & 9.09    & 0.03653    & 1.161    & E/cD    & Interacting/merging with bright galaxy and fainter one  \\[0.5ex]   
--    & 2214    & 321.599487    & 10.777511    & 0.095    & 741    & 0.818    & 3.98    & 0.02260    & 1.199    & E    &     \\[0.5ex]   
2096    & 2109    & 359.836166    & 14.670211    & 0.093    & 786    & 1.161    & 6.56    & 0.03572    & 1.242    & cD/E    &     \\[0.5ex]   
2085    & 2085    & 334.197449    & $-$9.724778    & 0.094    & 806    & 0.779    & 3.43    & 0.02861    & 1.348    & cD    &     \\[0.5ex]   
2027    & 2035    & 4.177309    & $-$0.445436    & 0.065    & 1084    & 1.436    & 8.89    & 0.02417    & 1.168    & cD    & Several companions  \\[0.5ex]   
--    & 3084    & 118.360820    & 29.359459    & 0.061    & 781    & 1.584    & 3.95    & 0.06632    & 1.382    & cD    & Several faint and bright companions  \\[0.5ex]   
--    & 3347    & 119.679733    & 30.773809    & 0.076    & 902    & 1.354    & 6.04    & 0.01470    & 1.019    & E/cD    &     \\[0.5ex]   
--    & 1283    & 125.745443    & 4.299105    & 0.095    & 754    & 2.747    & 10.47    & 0.04483    & 1.094    & cD    & Several faint-ish companions  \\[0.5ex]   
--    & 1039    & 186.878093    & 8.824560    & 0.090    & 846    & 1.962    & 6.94    & 0.06100    & 1.965    & cD    & Clear halo, bright companion (dumbbell galaxy)  \\[0.5ex]   
 
\hline\hline  
\end{tabular}
\caption{Properties of the BCG sample.  Columns (1) and (2)  provide galaxy identifications, where ID2 is the SDSS-C4 number $<$SDSS-C4 NNNN$>$ and ID3 is the SDSS C4$\_$2003 number, $<$SDSS-C4-DR3 NNNN$>$, as given in Simbad \citep{Linden07}. Columns (3) and (4) give the right ascension and declination in degrees. Column (5) gives the redshift and column (6) the velocity dispersion of the cluster. Columns (7), (8), (9) and (10) contain the effective radius, \sersic\ index, residual flux fraction and reduced $\chi^2$ derived from the single \sersic\ fits (see text for details). Column (11) gives the visual morphological classification of the BCGs. Column (12) contains some comments from the classifier. }
\end{table}
\label{lastpage}
\end{landscape}

\end{document}